\documentclass[12pt]{article}
\pdfoutput=1

\usepackage[a4paper,text={16.8cm,22.4cm}]{geometry}
\usepackage{amsmath,amsfonts,braket,slashed,amssymb,bm,psfrag,graphicx,color,dsfont,euscript,ctable}
\usepackage[small,labelfont=bf]{caption}

\RequirePackage[sort&compress,square,comma,numbers]{natbib}
\allowdisplaybreaks
\newcommand{\A}{{\EuScript A}}
\newcommand{\B}{{\EuScript B}}
\newcommand{\G}{{\EuScript G}}
\newcommand{\nbsl}{\rlap{\hspace{0.25mm}/}{\bar n}}
\newcommand{\vsl}{\rlap{\hspace{0.25mm}/}{v}}
\newcommand{\Asl}{\rlap{\hspace{0.6mm}/}{\A}}
\newcommand{\delsl}{\rlap{\hspace{0.2mm}/}{\partial}}
\newcommand{\nb}{{\bar n}}
\newcommand{\spac}{{\hspace{0.3mm}}}
\newcommand{\colored}[1]{{\color{gray}{#1}}}
  
\begin{document}

\begin{titlepage}

\begin{flushright}
\normalsize
MITP/20-053\\
TUM-HEP-1290/20\\
ZH-TH-39/20\\
November 16, 2020
\end{flushright}

\vspace{1.0cm}
\begin{center}
\Large\bf\boldmath
Effective Field Theory for Heavy Vector Resonances Coupled to the Standard Model
\end{center}

\vspace{0.5cm}
\begin{center}
Mathias Heiles$^a$, Matthias K\"onig$^{b,c}$ and Matthias Neubert$^{a,c,d}$\\
\vspace{0.7cm} 
{\sl ${}^a$PRISMA Cluster of Excellence \& Mainz Institute for Theoretical Physics\\
Johannes Gutenberg University, 55099 Mainz, Germany\\[3mm]
${}^b$Physik Department T31, Technische Universität M\"unchen\\
James-Frank-Straße 1, 85748 Garching, Germany\\[3mm]
${}^c$Physik-Institut, Universit\"at Z\"urich, CH-8057, Switzerland\\[3mm]
${}^d$Department of Physics \& LEPP, Cornell University, Ithaca, NY 14853, U.S.A.}
\end{center}

\vspace{0.8cm}
\begin{abstract}
We construct an effective field theory describing the decays of a heavy vector resonance $V$ into Standard Model particles. The effective theory is built using an extension of Soft-Collinear Effective Theory called SCET$_{\rm BSM}$, which provides a rigorous framework for parameterizing decay matrix elements with manifest power counting in the ratio of the electroweak scale and the mass of the resonance, $\lambda\sim v/m_V$. Using the renormalization-group evolution of the couplings in the effective Lagrangian, large logarithms associated with this scale ratio can be resummed to all orders. We consider in detail the two-body decays of a heavy $Z'$ boson and of a Kaluza-Klein gluon at leading and subleading order in $\lambda$. We illustrate the matching onto SCET$_{\rm BSM}$ with a concrete example of a UV-complete new-physics model.
\end{abstract}

\end{titlepage}

\tableofcontents

\section{Introduction}

After the discovery of a new heavy resonance, studying its couplings to the particles of the Standard Model (SM) will be of utmost importance. This can be achieved, in particular, by investigating on-shell decays of the resonance into SM final states. In order to separate the different energy scales relevant for these decay processes, the method of effective field theory (EFT) offers a valuable tool. Scale separation is desirable for two reasons: First, by separating the new physics from SM physics, it allows for a model-independent description that can be adapted to any specific new-physics scenario with only little effort. The effective theory is constructed as a systematic expansion in the small parameter $\lambda\sim v/\Lambda$, where we assume a large hierarchy between the electroweak scale $v\approx 246$\,GeV and the new-physics scale $\Lambda$. By using an EFT approach the processes of interest can be described without knowledge of the full underlying theory. Second, by constructing the appropriate operator basis the power counting is manifested at the level of the Lagrangian, while at the same time keeping large logarithms of the scale ratio under control.

The appropriate EFT for low-energy physics is an extension of the SM called SMEFT, in which the effective operators contain SM fields only, and new physics enters through the Wilson coefficients of these operators. However, this approach cannot provide a meaningful description of the on-shell decays of a new heavy resonances. The scale separation in this case is between the energy injected into the final-state particles and their mass, $E_i\sim m_V\gg m_i$, where $i$ labels the final-state particles. The appropriate framework for this situation is Soft-Collinear Effective Theory (SCET) \cite{Bauer:2000yr,Bauer:2001yt,Bauer:2002nz,Beneke:2002ph}, an effective theory for highly energetic light particles. In SCET every SM field is split up into momentum modes that are either collinear with an energetic final-state particle or soft. Modes with large virtualities are integrated out in the Wilsonian way. The effective theory is built from gauge-invariant building blocks, which are comprised of the soft and collinear fields dressed with Wilson lines.

In Refs.~\cite{Alte:2018nbn,Alte:2019iug}, the original formulation of SCET has been extended to an effective theory called SCET$_{\rm BSM}$, which describes light SM fields coupled to a field describing a hypothetical new heavy resonance with mass much above the electroweak scale. In these papers the operator basis was constructed up to third (and partially fourth) order in the expansion parameter $\lambda$ for the case of a scalar resonance $S$ transforming as a singlet under the SM gauge group. In the present work we extend the framework to the case of heavy vector resonances, while at the same time allowing for the resonance to carry SM charges. An additional subtlety in this case arises from the fact that theories with massive vector fields are {\it a priori\/} not renormalizable. The mass of the resonance must either be acquired via a Higgs-like mechanism, or the resonance must arise as a composite state from a confining strong interaction. Neither of these mechanisms are visible to the effective theory. Hence, the vector resonance needs to be described in a fashion analogous to the $b$ quark in Heavy-Quark Effective Theory (HQET) \cite{Eichten:1989zv,Georgi:1990um,Neubert:1993mb}, in which only the soft fluctuations of the heavy field remain as dynamical degrees of freedom. In this way, the effective Lagrangian is renormalized even though the vector resonance can appear as a (soft) virtual particle in loop diagrams. For most of this work we focus on the case of a vector resonance transforming as a SM singlet, which is commonly referred to as a $Z'$ boson. In this case no loop-amplitudes with virtual heavy vector resonances exist. Even then, however, introducing an effective field for the heavy vector meson is needed to restrict the operator basis and keep the EFT power counting consistent, because different components of the vector field obey different scaling rules in the EFT.

Massive, gauge-singlet vector bosons are common ingredients of many theories beyond the SM, typically appearing as gauge bosons of larger symmetry groups that are broken to the SM one. As such they appear frequently in models with a gauged $B-L$ charge, models of compositeness or theories with extra dimensions (see \cite{Langacker:2008yv} for a comprehensive review). In the context of the persisting $B$-physics anomalies, they have raised interest as potential mediators of the $b\to s\spac\ell^+\ell^-$ anomalies \cite{Buras:2013qja,Altmannshofer:2014cfa,Calibbi:2019lvs}, and they are part of the new-physics spectrum of more comprehensive models offering combined explanations of the anomalies in $b\to s\ell^+\ell^-$ and $b\to c\ell^-\bar\nu$ transitions \cite{DiLuzio:2017vat,Bordone:2017bld}. Even though the main mediators in the latter models are vector leptoquarks, it can be shown that a $Z'$ boson cannot be avoided in such constructions \cite{Baker:2019sli}. The framework outlined in this paper can be used to describe the decays of heavy vector resonances into SM fields, including the sizable resummation effects affecting the decay rates, in an economic and systematic way.

This paper is organized as follows: In Section~\ref{sec:SCET} we begin with a brief review of the necessary ingredients of SCET, followed by a discussion of the construction of the Heavy-Vector Effective Theory (HVET) needed to describe the heavy resonance in the effective Lagrangian. For the most interesting case of a gauge-singlet resonance $Z'$, we then discuss in detail the construction of the operator basis for two-prong decays at leading and subleading order in power counting. In Section~\ref{sec:matrixelem} we derive the corresponding decay amplitudes at tree level. The renormalization-group (RG) running of the matching coefficients and the resummation of large logarithms is discussed in Section~\ref{sec:RGE}. In Section~\ref{sec:nonsinglet} we generalize our framework to vector resonances that are charged under the SM gauge group. Finally, we illustrate our approach with the example of a specific UV completion in Section~\ref{sec:matching}. Our conclusions are given in Section~\ref{sec:concl}.

\renewcommand{\theequation}{2.\arabic{equation}}
\setcounter{equation}{0}

\section{Construction of the effective theory}
\label{sec:SCET}

\subsection{Basic elements of SCET}

We now give a brief overview of the necessary ingredients of SCET relevant to this paper, referring the reader to \cite{Alte:2018nbn} for a more detailed discussion. SCET$_{\rm BSM}$ is an EFT for energetic light SM particles coupled to a new heavy particle, which serves as a source of large energy. All operators in this EFT must be invariant under the (unbroken) SM gauge group. For each direction of large momentum flow, one specifies a set of light-like reference vectors $n_i$ and $\nb_i$, satisfying $n_i^2=\nb_i^2=0$ and $n_i\cdot\nb_i=2$. In the present work, we limit our attention to two-prong final states, which in the rest frame of the decaying resonance are always in a back-to-back configuration. We can thus choose a single set of reference vectors
\begin{align}
   n^\mu = (1,0,0,1) \,, \qquad
   \nb^\mu = (1,0,0,-1) \,,
\end{align}
each denoting a direction of large momentum flow. Any 4-momentum can then be decomposed as
\begin{align}\label{eq2}
   k^\mu = \frac{n^\mu}{2}\,\nb\cdot k + \frac{\nb^\mu}{2}\,n\cdot k + k_{\perp}^\mu \,.
\end{align}
In a similar manner, the metric tensor can be decomposed as 
\begin{align}
   g_{\mu\nu} = g_{\mu\nu}^\perp + \frac{n_\mu\nb_\nu+\nb_\mu n_\nu}{2} \,.
\end{align}
Lorentz indices in the transverse plane orthogonal to the reference vectors $n^\mu$ and $\nb^\mu$ can be contracted with the help of the symmetric and antisymmetric two-index tensors $g_{\mu\nu}^\perp$ and 
\begin{align}
   \epsilon_{\mu\nu}^\perp = \epsilon_{\mu\nu\rho\sigma}\,\frac{n^\rho\spac\nb^\sigma}{2} \,.
\end{align}

In general, the effective theory for two-prong decays contains fields whose momenta obey the scaling laws
\begin{align}\label{scalings}
\begin{aligned}
  (\nb\cdot k,n\cdot k,k_\perp) 
  &\sim E \left( 1,\lambda^2,\lambda \right) ; & & \text{``collinear''} \\
  (\nb\cdot k,n\cdot k,k_\perp) 
  &\sim E \left( \lambda^2,1,\lambda \right) ; & & \text{``anti-collinear''} \\
  (\nb\cdot k,n\cdot k,k_\perp) 
  &\sim E \left( \lambda,\lambda,\lambda \right) ; & & \text{``soft''} \\
  (\nb\cdot k,n\cdot k,k_\perp) 
  &\sim E \left( \lambda^2,\lambda^2,\lambda^2 \right) ; & \quad & \text{``ultra-soft''}
\end{aligned}
\end{align}
where $E\sim m_V$ is the characteristic decay energy, which scales with the mass of the decaying resonance. Note that ultra-soft modes can only arise for massless fields in the low-energy theory. Since some of the low-energy degrees of freedom can have large momentum components (of order $E$), derivatives acting on (anti-)collinear fields are not necessarily power-suppressed, and hence the most general effective Lagrangian contains an infinite number of operators with arbitrary powers of large derivatives. They can be traded for non-localities of the composite operators and their Wilson coefficients along light-like directions, over which the Lagrangian is then integrated. In momentum space, this important feature of SCET results in a dependence of the Wilson coefficients on the mass of the decaying resonance in addition to the masses of other heavy particles that have been integrated out. All remaining derivatives on SCET fields correspond to small momentum components and give rise to power-suppressed contributions.

In the presence of the above-mentioned non-local operator products, gauge invariance is restored by means of collinear and anti-collinear Wilson lines, defined as
\begin{align}
   W_n^{(A)}(x) 
   = P \exp\left[ ig_A\,t_A^a \int_{-\infty}^0 ds\,\nb\cdot A_n^a(x+s\nb) \right] ,
\end{align}
and similarly for $W^{(A)}_\nb(x)$. Here $g_A$ denotes the gauge coupling associated with the gauge field $A$, and $t_A^a$ is the group generator in the representation of the field on which the Wilson line acts. For the hypercharge Wilson line $W_n^{(B)}$ the generator is replaced by the hypercharge operator. The object $P$ is the path-ordering symbol. Wilson lines often play no role in tree-level calculations in SCET, because they correspond to longitudinally polarized gauge fields, which are unphysical for massless fields. For final states containing electroweak bosons, however, the Wilson lines do generate physical final states, as we will see later.

The relevant SCET operators consist of products of ``building blocks'' comprised of the fields dressed with Wilson lines~\cite{Bauer:2002nz,Hill:2002vw}. For the Higgs doublet, the $n$-collinear building block is
\begin{align}
   \Phi_n(x) = W_n^\dagger(x)\,\phi_n(x) \,, \qquad
   W_n(x)\equiv W_n^{(W)}(x)\,W_n^{(B)}(x) \,.
\end{align}
Here $\phi_n(x)$ refers to a projection of the quantum field $\phi(x)$ for the SM Higgs doublet onto its Fourier modes carrying a collinear momentum with scaling as shown in (\ref{scalings}). The building block for an $n$-collinear fermion is given by
\begin{align}\label{eq:fermionBlock}
   \Psi_n(x) = P_n\,W_n^\dagger(x)\,\psi_n(x) 
   = \frac{\slashed n\slashed\nb}{4}\,W_n^\dagger(x)\,\psi_n(x)\,, 
\end{align}
where $W_n$ is the appropriate product of Wilson lines depending on the charges of the fermion under the SM gauge group. The projection operator $P_n$ singles out the large components of the spinor of a highly energetic fermion. Finally, for an $n$-collinear gauge field the building block is defined as
\begin{align}
\begin{aligned}
   \A_n^\mu(x)\equiv \A_n^{\mu,a}(x)\,t_A^a
   &= W_n^{(A)\dagger}(x)\,\big[ iD^\mu_n\, W_n^{(A)}(x) \big] \\
   &= g_A \int_{-\infty}^0 ds\,\nb_\nu\,\big[ W_n^{(A)\dagger} F_n^{\nu\mu}\,W_n^{(A)} 
    \big](x+s\nb) \,.
\end{aligned}
\end{align}
Here $D_n^\mu$ and $F_n^{\nu\mu}$ are the covariant derivative and field-strength tensor associated with the collinear gauge field $A_n^\mu$. Written in this form the gauge fields contain the generators $t_A^a$ of the gauge group, and in the case of the $U(1)_Y$ gauge field we define $\B_n^\mu(x)\equiv \B_n^{\mu,a}(x)\,Y$, where $a=1$ and $Y$ is the hypercharge operator. By definition, the $n$-collinear building blocks satisfy the constraints 
\begin{align}\label{constraints}
   \slashed n\,\psi_n(x) = 0 \,, \qquad \nb\cdot\A_n(x) = 0 \,.
\end{align}

An important fact is that one can associate a consistent power counting in $\lambda$ with the effective fields defined in SCET, finding
\begin{align}\label{fieldscalings}
   \Phi_n \sim \lambda \,, \qquad
   \Psi_n \sim \lambda \,,  \qquad 
   \A_n^\perp \sim \lambda \,, \qquad
   n\cdot\A_n \sim \lambda^2 \,.
\end{align}
Importantly, the large component $\bar n\cdot\A_n$, which would scale like $\lambda^0$, vanishes by virtue of the constraint (\ref{constraints}). The anti-collinear building blocks, defined with $n$ and $\nb$ exchanged everywhere, obey the same scaling rules. In addition, one can define soft and ultra-soft fields in the EFT, which obey the scaling laws 
\begin{align}
   \phi_s \sim \lambda \,, \qquad
   \psi_s \sim \lambda^{3/2} \,,  \qquad 
   A_s^\mu \sim \lambda \,,
\end{align}
and
\begin{align}
   \phi_{us} \sim \lambda^2 \,, \qquad
   \psi_{us} \sim \lambda^3 \,,  \qquad 
   A_{us}^\mu \sim \lambda^2 \,.
\end{align}
It follows that adding more fields to an operator always increases its scaling with $\lambda$, giving rise to higher power suppression in the ratio of the electroweak scale to the new-physics scale. This important fact ensures that, as long as one is interested in the leading and subleading terms in the power expansion, a relatively small number of operators contributes. Operators in the EFT can also contain derivatives $\partial^\mu$ acting on the fields, but only those derivatives corresponding to small momentum components can appear, which again gives rise to a suppression in powers of $\lambda$.

Besides the usual momentum modes in SCET, one also needs to introduce a static mode $\Phi_0\sim \lambda$ carrying no four-momentum in order to encode the effects of electroweak symmetry breaking (EWSB) \cite{Alte:2018nbn}. In the broken phase, this mode will be replaced by the Higgs vacuum expectation value (VEV), i.e.\
\begin{align}
   \Phi_0 \stackrel{\mathrm{EWSB}}{\to} \frac{1}{\sqrt{2}}
    \left( \begin{array}{c} 0 \\ v \end{array} \right) .
\end{align}

\subsection{Heavy-Vector Effective Theory}

We will describe the heavy vector resonance $V$, which in general can be charged or uncharged under the SM gauge group, in terms of an effective vector field $V_v^\mu(x)$ subject to the constraint $v\cdot V_v(x)=0$. Here $v^\mu$ with $v^2=1$ is the 4-velocity of the on-shell resonance (not to be confused with the Higgs VEV), and in the rest frame $v^\mu=(1,\bm{0})$. For the case where the resonance is charged under the SM, it can interact in the EFT with soft or ultra-soft SM gauge bosons, which can take its momentum $p^\mu$ off-shell by a small amount, such that
\begin{align}\label{eq:momentumDecompose}
   p^\mu = m_V\spac v^\mu + k^\mu \,,
\end{align}
with $k^\mu/m_V={\cal O}(\lambda)$ or ${\cal O}(\lambda^2)$. While an on-shell vector boson satisfies the constraint $v\cdot\varepsilon_V=0$ trivially, the time-like mode $v\cdot V$ might still propagate in EFT loop diagrams, but it cannot do so as a hard degree of freedom, since hard quantum fluctuations are -- by definition -- already integrated out. In order to separate the relevant scales in a consistent way, one needs to decompose the resonance into an on-shell component and soft fluctuations of small virtualities, using a construction analogous to HQET \cite{Eichten:1989zv,Georgi:1990um,Neubert:1993mb}. 

Another potential problem in this context is that theories with massive vector fields are, in general, not renormalizable. The problem can be seen by considering the propagator
\begin{align}\label{eq:procaPropagator}
   \Pi_V^{\mu\nu}(p) = \frac{i}{p^2-m_V^2} \left( - g^{\mu\nu} + \frac{p^\mu\spac p^\nu}{m_V^2} \right) , \end{align}
which gives rise to highly divergent terms in loop calculations. For example, when computing the self-energy of the vector field, the second term leads to divergences that cannot be absorbed by counterterms of the form $i\delta_V (p^2 g^{\mu\nu}-p^\mu p^\nu)+i\delta_m g^{\mu\nu}$, because they are proportional to higher powers of $p^2$. We will now show that the problematic terms originate from modes of the vector field that live at or above the hard scale. Once these modes are integrated out in the construction of the HVET, the resulting EFT is renormalizable in the usual sense.

For concreteness, we perform the construction of the HVET for the case of a (real) massive vector field $V_\mu^a$ transforming in the adjoint representation of $SU(3)_c$. Such a particle appears, e.g., in the form of a Kaluza-Klein gluon in extensions of the SM with a compact extra dimension. The corresponding Lagrangian reads
\begin{align}\label{LKKgluon}
   \mathcal L_V = - \frac{1}{4}\,(D_\mu V_\nu - D_\nu V_\mu)^a\,(D^\mu V^\nu - D^\nu V^\mu)^a
    + \frac{m_V^2}{2}\,V_\mu^a\spac V^{\mu,a} \,,
\end{align}
where $(D_\mu V_\nu)^a=\partial_\mu V_\nu^a+g_s f^{abc}\spac G_\mu^b\spac V_\nu^c$ is the covariant derivative of the vector field, and $G_\mu^a$ denotes the ordinary gluon field. We now introduce the projection operators
\begin{align}
   P_\parallel^{\mu\nu} = v^\mu\spac v^\nu \,, \qquad
   P_{\perp_v}^{\mu\nu} = g^{\mu\nu} - v^\mu\spac v^\nu \,,
\end{align}
where the symbol $\perp_v$ means the projection of a vector onto its components perpendicular to the 4-vector $v^\mu$, which is different from the $\perp$ symbol defined in (\ref{eq2}). We can then split up the vector field $V_\mu^a$ into longitudinal and transverse polarization states, i.e.\
\begin{align}
   V^{\mu,a} = P_\parallel^{\mu\nu}\,V_\nu^a + P_{\perp_v}^{\mu\nu}\,V_\nu^a 
   \equiv V_\parallel^{\mu,a} + V_{\perp_v}^{\mu,a} \,.
\end{align}
Studying the propagator of these fields we find that the two modes have different power counting, namely
\begin{align}
   \braket{0|\,T\{V_i^{\mu,a}(x),V_i^{\nu,b}(0)\}|0}
   = \int\frac{d^4p}{(2\pi)^4}\,\frac{i\delta^{ab}\spac e^{-ip\cdot x}}{p^2-m_V^2} 
    \left\{ \begin{array}{cl}
     - \frac{2v\cdot k}{m_V}\,v^\mu\spac v^\nu + \mathcal{O}(k^2/m_V^2) \,;~ & i=\parallel \,, \\
      g^{\mu\nu} - v^\mu\spac v^\nu + \mathcal{O}(k^2/m^2_V) \,;~ & i=\perp_v .
    \end{array} \right.
\end{align}
We observe that the component $V_\parallel^{\mu,a}$ is suppressed, with respect to $V_{\perp_v}^{\mu,a}$, by one power of $k/m_V$. It is this small component that is integrated out in the construction of HVET.

More accurately, we define effective field operators via
\begin{align}\label{HVETfields}
\begin{aligned}
   \big[ V^{\mu,a}(x) \big]_\text{in} 
   &\equiv e^{-im_V v\cdot x}\,V_\text{eff}^{\mu,a}(x)  
    = e^{-im_V v\cdot x}\,\big[ V_v^{\mu,a}(x) + v^\mu\,\mathbb{V}_v^a(x) \big] \,, \\
   \big[ V^{\mu,a}(x) \big]_\text{out} 
   &\equiv e^{im_V v\cdot x}\,V_\text{eff}^{\mu,a\spac\dagger}(x)
    = e^{im_V v\cdot x}\,\big[ V_v^{\mu,a\spac\dagger}(x) + v^\mu\,\mathbb{V}_v^{a\spac\dagger}(x) \big] \,, 
\end{aligned}
\end{align}
where
\begin{align}
   V_v^{\mu,a}(x) = V_{\text{eff},\perp_v}^{\mu,a}(x) \,, \qquad
   \mathbb{V}_v^a(x) = v_\mu V_{\text{eff},\parallel}^{\mu,a}(x) \,,
\end{align}
and 
\begin{align}\label{vVzero}
   v_\mu V_v^{\mu,a}(x) = 0 \,.
\end{align}
The label ``$v$'' indicates that these are effective fields describing initial-state heavy vector particles with 4-velocity $v^\mu$. Because of the phase factors $e^{\mp im_V v\cdot x}$ pulled out, these fields carry the residual momentum $k^\mu$ defined in (\ref{eq:momentumDecompose}). The subscript ``in'' (``out'') in the first (second) relation of (\ref{HVETfields}) means that one must keep only the terms involving annihilation (creation) operators in the Fourier representation of the quantum field $V^{\mu,a}(x)=[V^{\mu,a}(x)]_\text{in}+[V^{\mu,a}(x)]_\text{out}$.

Inserting the decompositions (\ref{HVETfields}) into the Lagrangian (\ref{LKKgluon}) of the vector resonance, and keeping only the soft interaction terms, in which the phase factors $e^{\pm im_V v\cdot x}$ cancel out, one obtains
\begin{align}
   \mathcal L_V^{\rm eff} 
   = -g_{\mu\nu}\,m_V \left[ V_v^{\mu,a\spac\dagger}\spac(iv\cdot D\spac V_v^\nu)^a 
    - \mathbb{V}_v^{a\spac\dagger}\spac iD^\mu\spac V_v^{\nu,a} + \text{h.c.} \right]
    + m_V^2\,\mathbb{V}_v^{a\spac\dagger}\spac\mathbb{V}_v^a 
    + {\cal O}(m_V^0) \,,
\end{align}
where the terms of ${\cal O}(m_V^0)$ contain two derivatives on the fields and hence correspond to terms of quadratic order in the small residual momentum. Solving the classical equations of motion for the field $\mathbb{V}_v^a$ yields
\begin{align}
   \mathbb{V}_v^a = - \frac{1}{m_V}\,iD_\mu\spac V_v^{\mu,a} + {\cal O}\bigg( \frac{1}{m_V^2} \bigg) \,.
\end{align}
Inserting this solution back into the Lagrangian leads to
\begin{align}\label{LeffHVET}
   \mathcal L_\mathrm{HVET} 
   = 2 m_V \left[ \big( -P_{\perp_v}^{\mu\nu} \big)\,V_v^{\mu,a\spac\dagger}\spac(iv\cdot D\spac V_v^\nu)^a 
    + {\cal O}\bigg( \frac{1}{m_V} \bigg) \right] .
\end{align}
The above result is analogous to the HQET Lagrangian, which is indeed a consequence of heavy-quark spin symmetry \cite{Neubert:1993mb}. We have refrained from performing a rescaling of the field $V_v$ by $1/\sqrt{2m_V}$ (which would remove the prefactor $2m_V$) for the sake of this field retaining the canonical mass dimension of a vector field. The Feynman propagator for the vector field in HVET takes the form
\begin{align}
   \Pi_V^{\mu\nu}(k) 
   = \frac{-i\spac P_{\perp_v}^{\mu\nu}}{2m_V\spac(v\cdot k+i\epsilon)} \,,
\end{align}
which does not possess the problematic behavior for large momenta that we saw in \eqref{eq:procaPropagator}. Consequently, the effective theory is renormalizable independently of the origin of the vector-boson mass. This is indeed to be expected, since the EFT does not know about the details of its UV completion. The Feynman rule for the coupling of a colored vector field to a gluon with color index $a$ and Lorentz index $\alpha$ is 
\begin{align}
   2m_V\spac\big( -P_{\perp_v}^{\mu\nu} \big)\,g_s\spac f^{abc}\spac v^\alpha \,,
\end{align}
where $\mu$, $b$ ($\nu$, $c$) are the Lorentz and color indices of the outgoing (incoming) vector field.

For most of our discussion we will consider a gauge-singlet vector boson $Z'$, for which the covariant derivative in (\ref{LeffHVET}) must be replaced by an ordinary derivative and the color index $a$ must be dropped. At leading order in $1/m_V$ the effective Lagrangian is that of a free particle without interactions. The on-shell momentum of the $Z'$ bosons can be written as $p_{Z'}^\mu=m_V\spac v^\mu$ (without a residual momentum $k^\mu$), where without loss of generality we can choose $v_\perp^\mu=0$. The most important property of the HVET field $Z_v^{\prime\mu}$ in this case is that it satisfies the constraint (\ref{vVzero}). We will return to the case of a color-octet vector resonance in Section~\ref{sec:nonsinglet}.

\subsection{Operator basis at leading order}
\label{subsec:2.3}

We now proceed to construct the operator basis of the effective Lagrangian at leading and subleading order in $\lambda=v/m_V$, considering the case of a gauge-singlet vector field $Z'$ for concreteness. We will derive the effective Lagrangian for two-prong decays of this new resonance. The relevant operators must contain at least one $n$-collinear and one $\nb$-collinear field, and since according to (\ref{fieldscalings}) each collinear field scales with at least one power of $\lambda$, the leading operators start at $\mathcal O(\lambda^2)$. Note that all operators must be invariant under the exchange of the reference vectors $n$ and $\nb$. Gauge and Lorentz invariance then allow for three options, namely operators containing a $Z_v^{\prime \mu}$ field along with a pair of fermions, a pair of Higgs fields or a pair of transverse gauge fields. In each case the Lorentz index of the $Z'$ field must be contracted with some 4-vector index. For the case of fermions, the only possible structure is
\begin{align}\label{Opsipsi}
   O_{\psi\psi}^{ij} 
   = Z_{v\mu}' \big( \bar\Psi^i_n\gamma_\perp^\mu\Psi^j_\nb
    + \bar\Psi^i_\nb\gamma_\perp^\mu\Psi^j_n \big) \,, 
\end{align}
where $\gamma_\perp^\mu$ is defined as in (\ref{eq2}). The effective fermion field $\Psi$ can represent any one of the chiral SM fermion multiplets $Q_L,d_R,u_R,L_L,e_R$, where $i,j=1,2,3$ are generation indices. The leading-order couplings of the $Z'$ boson to fermions must necessarily involve fermion fields with equal chiralities, because fermion bilinears made out of opposite-chirality fields need an insertion of the Higgs doublet for gauge invariance. Such operators are thus of higher order in power counting. One may wonder if there exists a second operator of the type shown in (\ref{Opsipsi}), in which the field $Z_v^{\prime\mu}$ and the Dirac matrix $\gamma_\perp^\nu$ are contracted using the antisymmetric $\epsilon_{\mu\nu}^\perp$ symbol. However, because of the identity \cite{Hill:2004if}
\begin{align}\label{Hillrela}
   P_n^\dagger\,\epsilon_{\mu\nu}^\perp \gamma_\perp^\nu P_\nb
   = i P_n^\dagger\,\gamma_\mu^\perp\gamma_5 P_\nb
\end{align}
this does not give rise to a new structure, as $\gamma_5$ can be replaced by one of its eigenvalues $\pm 1$ when acting on the chiral fermion fields of the SM.

For the case of the $Z'$ coupling to two Higgs fields, we need to construct a 4-vector that can be contracted with the effective field $Z_v^{\prime \mu}$. Since any derivative acting on a field in the EFT is power-suppressed in $\lambda$, we must build this 4-vector out of the reference vectors $v$, $n$ and $\nb$. Owing to the condition $v\cdot Z_v'=0$, the only possible choice is
\begin{align}\label{Pidef}
   \Pi^\mu = \frac{(v\cdot\nb)\,n^\mu - (v\cdot n)\,\nb^\mu}{2} \,,
\end{align}
which in the rest frame of the $Z'$ boson and with our standard choice of reference vectors evaluates to $\Pi^\mu=(0,0,0,1)$. Note that this object is odd under the exchange of $n$ and $\nb$. It follows that the relevant operator is
\begin{align}\label{Ophiphi}
   O_{\phi\phi} 
   = m_{Z'}\,\Pi\cdot Z_v'\,\big( \Phi_n^\dagger\Phi_\nb - \Phi_\nb^\dagger\Phi_n \big) \,.
\end{align}
The factor $m_{Z'}$ is inserted here to ensure that the Wilson coefficient of this operator is dimensionless. Despite appearance, the operator $O_{\phi\phi}$ is hermitian. The argument showing this is somewhat subtle, so let us explain it in detail. Consider the hermitian current operator $\phi^\dagger i\!\overleftrightarrow{D}^{\!\mu}\phi$, which may arise in some UV completion of the SM. When mapped onto SCET, the field $\phi$ gets replaced by the building blocks $\phi\to\Phi_n+\Phi_\nb+\dots$ up to a possible soft contribution, and the leading-order contributions (in powers of $\lambda$) arise from the large components of the derivatives acting on the collinear fields. One obtains
\begin{align}
   \phi^\dagger i\!\overleftrightarrow{D}^{\!\mu}\phi
   \to \Phi_n^\dagger \left( \frac{\nb^\mu}{2}\,in\cdot\partial 
    - \frac{n^\mu}{2}\,i\nb\cdot\!\overleftarrow{\partial}\! \right) \Phi_\nb
    + \Phi_\nb^\dagger \left( \frac{n^\mu}{2}\,i\nb\cdot\partial 
    - \frac{\nb^\mu}{2}\,in\cdot\!\overleftarrow{\partial}\! \right) \Phi_n + \dots \,,
\end{align}
up to power-suppressed terms. If the fields carry outgoing momenta $p_n^\alpha$ and $p_\nb^\alpha$ satisfying $p_n^\alpha+p_\nb^\alpha=m_{Z'}v^\alpha$, like in our case, then the leading terms reduce to
\begin{align}
   \phi^\dagger i\!\overleftrightarrow{D}^{\!\mu}\phi
   \to m_{Z'} \Pi^\mu \big( \Phi_n^\dagger\Phi_\nb - \Phi_\nb^\dagger\Phi_n \big) + \dots \,.
\end{align}
The point is that under a hermitian conjugation the effective field for the heavy vector meson $Z_v^{\prime\mu}$, which only contains an annihilation operator, is transformed into the conjugate field $Z_v^{\dagger\prime\mu}$, which contains a creation operator. In other words, the $Z'$ boson is moved from the initial to the final state. One needs to perform a crossing transformation to move this particle back into the initial state, and in this processes one must replace $v^\mu\to -v^\mu$. It follows that under hermitian conjugation the quantity $\Pi^\mu$ defined in (\ref{Pidef}) changes sign, and hence the operator $O_{\phi\phi}$ is indeed hermitian.

Finally, for the case where the $Z'$ field couples to two transverse gauge fields, the transverse Lorentz indices of the gauge fields must be contracted with each other, which only leaves the possibility to contract the $Z'$ field with the object $\Pi^\mu$. The fact that the gauge fields are hermitian then implies that only the combination
\begin{align}\label{ZpAA}
   m_{Z'}\,\Pi\cdot Z_v'\,\epsilon_{\mu\nu}^\perp\,\A_n^{\perp\mu,a} \A_\nb^{\perp\nu,a}
\end{align}
could potentially be non-zero. However, it can easily be seen that the contraction $\epsilon_{\mu\nu}^\perp\,\A_n^{\perp\mu} \A_\nb^{\perp\nu}$ is even under the exchange of $n$ and $\nb$, and hence the above operator is not allowed.

This exhausts all options at $\mathcal{O}(\lambda^2)$. The most general effective Lagrangian at this order is therefore (a sum over repeated indices is implied)
\begin{align}\label{eq:LagLambda2}
   \mathcal{L}_\mathrm{eff}^{(2)}
   = C_{\phi\phi}(m_{Z'},\mu)\,O_{\phi\phi}(\mu)
    + \sum_\psi C_{\psi\psi}^{ij}(m_{Z'},\mu)\,O^{ij}_{\psi\psi}(\mu) 
\end{align}
where the sum in the second term runs over all SM fermion multiplets. The Wilson coefficient $C_{\phi\phi}$ is real, while the coefficients $C_{\psi\psi}^{ij}$ form the entries of $3\times 3$ hermitian matrices. If these matrices are complex, then the fermion operators can mediate CP-violating interactions, in analogy with the CKM matrix in the SM. As we will see later, after EWSB the first operator generates the two-body decays $Z'\to hZ$ and $Z'\to W^+W^-$, but not $Z'\to ZZ$, $Z'\to Z\gamma$ and $Z'\to h\gamma$. The decay $Z'\to hh$ is forbidden by angular momentum conservation, while $Z'\to\gamma\gamma$ and $Z'\to gg$ are forbidden by the Landau-Yang theorem \cite{Landau:1948kw,Yang:1950rg}.

\subsection{Operator basis at subleading order}

At ${\cal O}(\lambda^3)$ in SCET power counting a large number of operators contribute. A complete basis of these operators is given in Appendix~\ref{app:A}. It includes same-chirality fermion operators (like $O_{\psi\psi}$) containing an additional gauge boson, opposite-chirality fermion operators containing an additional Higgs doublet, di-Higgs operators (like $O_{\phi\phi}$) containing an additional gauge boson, and operators containing two or three gauge fields. Some of these operators contain a soft gauge or Higgs field. They are relevant for the description of decay processes in which one allows for soft radiation in addition to the two energetic final-state particles (or particle jets). In the operators containing three collinear fields, two of them belong to the same ($n$-collinear or $\nb$-collinear) jet. These operators are relevant for higher-order calculations, in which one allows for loops of collinear particles or collinear emissions inside the same jet. 

\begin{table}
\centering 
\begin{tabular}{|c||c|c|c|} 
\hline
Channel & $O_{\phi\phi}$ & $O_{B\phi}^{\parallel,\perp}$ & $O_{W\phi}^{\parallel,\perp}$ \\
\hline\hline 
$Z'\to W^+ W^-$ & $\lambda^2$ & & $\lambda^3$ \\ 
\hline
$Z'\to ZZ$ & & $\lambda^3$ & $\lambda^3$ \\ 
\hline
$Z'\to Z\gamma$ & & $\lambda^3$ & $\lambda^3$ \\ 
\hline
$Z'\to hZ$ & $\lambda^2$ & $\lambda^3$ & $\lambda^3$ \\ 
\hline
$Z'\to h\gamma$ & & $\lambda^3$ & $\lambda^3$ \\ 
\hline
\end{tabular}
\caption{Overview of the contributions of various operators to the di-boson decay amplitudes.}
\label{tab:operators}
\end{table}

In this work we are primarily interested in lowest-order predictions for the decay amplitudes, which we will then improve by resumming large logarithmic corrections arising in the RG evolution from the new-physics scale down to the electroweak scale. Of all the ${\cal O}(\lambda^3)$ operators listed in Appendix~\ref{app:A}, we then need to focus on the three operators  
\begin{align}\label{lam3ops}
\begin{aligned}
   O_{\psi_L\psi_R}^{ij} 
   &= \frac{\Pi\cdot Z_v'}{m_{Z'}}\,\big( \bar\Psi_{L,n}^i\Phi_0\Psi_{R,\nb}^j 
    - \bar\Psi_{L,\nb}^i\Phi_0\Psi_{R,n}^j \big) \,, \\[-1mm]
   O_{A\phi}^\parallel &= g_{\mu\nu}^\perp\,Z_v^{\prime\mu}\,\big( 
    \Phi_n^\dagger\,\A_\nb^{\perp\nu} \Phi_0
    + \Phi_\nb^\dagger\,\A_n^{\perp\nu} \Phi_0 \big) \,, \\[1mm]
   O_{A\phi}^\perp &= \epsilon_{\mu\nu}^\perp\,Z_v^{\prime\mu}\,\big( 
    \Phi_n^\dagger\,\A_\nb^{\perp\nu} \Phi_0
    - \Phi_\nb^\dagger\,\A_n^{\perp\nu} \Phi_0 \big) \,,
\end{aligned}
\end{align}
where $i,j$ are generation indices, and the gauge fields in the last two operators refer to either the hypercharge or the $SU(2)_L$ gauge group. Note that the transverse Levi-Civita symbol transforms with a sign change under $n\leftrightarrow\nb$, and hence the remaining terms in the operator $O_{A\phi}^\perp$ must also be odd under this exchange. The first operator above mediates decays of the $Z'$ boson into two fermions with opposite chiralities. After EWSB the remaining, purely bosonic operators mediate the decays $Z'\to ZZ$, $Z'\to Z\gamma$ and $Z'\to h\gamma$, which are not generated by ${\cal O}(\lambda^2)$ operators, and they also give power-suppressed contributions to the decays $Z'\to W^+ W^-$ and $Z'\to hZ$. We summarize the operators contributing to the various bosonic decay modes in Table~\ref{tab:operators}. The superscripts $\parallel$ and $\perp$ indicate that in the matrix elements of the operators the transverse polarization vectors of the $Z'$ boson and the relevant gauge boson appear in the combination $\bm{\varepsilon}_{Z'}\cdot\bm{\varepsilon}_A^*$ and $\bm{\varepsilon}_{Z'}\times\bm{\varepsilon}_A^*$, respectively (in 3-vector notation). We write the effective Lagrangian at order ${\cal O}(\lambda^3)$ as 
\begin{align}\label{eq:LagLambda3}
   \mathcal L_\mathrm{eff}^{(3)}
   = \sum_{\psi_L,\psi_R} C_{\psi_L\psi_R}^{ij}(m_{Z'},\mu)\,O_{\psi_L\psi_R}^{ij}(\mu) 
    + \sum_{A=B,W}\,\sum_{\sigma=\parallel,\perp} C_{A\phi}^\sigma(m_{Z'},\mu)\,O_{A\phi}^\sigma(\mu) 
    + \text{h.c.} + \dots \,,
\end{align}
where the dots refer to the many other operators listed in Appendix~\ref{app:A}, which do not contribute at tree level to the amplitudes we consider. In the first term $\psi_L$ and $\psi_R$ are summed over the left- and right-handed fermion multiplets of the SM, where left-handed quarks must be paired with right-handed quarks, and likewise for leptons. For the case where $\psi_R=u_R$ the replacement $\Phi\to\tilde\Phi$ with $\tilde\Phi_a=\epsilon_{ab}\,\Phi_b^*$ must be made to ensure gauge invariance. The Wilson coefficients in this Lagrangian are arbitrary complex quantities. 

At ${\cal O}(\lambda^3)$ in power counting, it is also possible to couple the $Z'$ boson to two gauge fields. The relevant operators read
\begin{align}\label{redundantops}
\begin{aligned}
   \tilde O_{AA}^\parallel &= i m_{Z'}\,g_{\mu\nu}^\perp\,Z_v^{\prime\mu} \left( 
    v\cdot\A_n^a\,\A_\nb^{\perp\nu,a} + v\cdot\A_\nb^a\,\A_n^{\perp\nu,a} \right) , \\
   \tilde O_{AA}^\perp &= i m_{Z'}\,\epsilon_{\mu\nu}^\perp\,Z_v^{\prime\mu} \left( 
    v\cdot\A_n^a\,\A_\nb^{\perp\nu,a} - v\cdot\A_\nb^a\,\A_n^{\perp\nu,a} \right) ,
\end{aligned}
\end{align}
where a sum over color indices $a$ is implied for the non-abelian gauge fields. In the products $v\cdot\A_n^a$ and $v\cdot\A_\nb^a$ only the power-suppressed components of the gauge fields of ${\cal O}(\lambda^2)$ contribute, e.g.\ $v\cdot\A_n^a=\frac{v\cdot\bar n}{2}\,n\cdot\A_n^a$. Our definitions of the operators $\tilde O_{AA}^{\parallel,\perp}$ contain a factor of $i$, such that these operators are hermitian (recall that $v^\mu\to-v^\mu$ under hermitian conjugation). We use a tilde to indicate that, consequently, the matrix elements of these operators are odd under a CP transformation. It is admissible to omit these two operators from our basis because they can be eliminated using the equations of motion  \cite{Marcantonini:2008qn}. This is discussed in more detail in Appendix~\ref{app:A}. We find that
\begin{align}\label{eq37}
   \tilde O_{AA}^\sigma = - i g_A^2\spac O_{A\phi}^\sigma + \text{h.c.} + \dots \,; 
    \quad \sigma=\,\,\parallel,\perp \,,
\end{align}
where $g_A=g$ or $g'$ for $A=W$ or $B$ denote the gauge couplings of $SU(2)_L$ and $U(1)_Y$, respectively, and the dots refer to other operators not contained in (\ref{redundantops}). As a consequence of these relations, we find that adding the redundant operators  (\ref{redundantops}) to our basis would simply redefine the imaginary parts of the Wilson coefficients $C_{A\phi}^\sigma$ according to
\begin{align}
   \Im m\,C_{W\phi}^\sigma \to \Im m\,C_{W\phi}^\sigma - g^2\spac\tilde C_{WW}^\sigma \,, \qquad
   \Im m\,C_{B\phi}^\sigma \to \Im m\,C_{B\phi}^\sigma - g^{\prime\,2}\spac\tilde C_{BB}^\sigma \,. 
\end{align}

\renewcommand{\theequation}{3.\arabic{equation}}
\setcounter{equation}{0}

\section{Matrix elements and decay rates}
\label{sec:matrixelem}

We now move on to present the two-body decay rates of a hypothetical heavy $Z'$ boson into pairs of SM particles derived at tree level from the effective SCET$_{\rm BSM}$ Lagrangian
\begin{align}
   \mathcal L_\mathrm{eff}
   = \mathcal L_\mathrm{eff}^{(2)} + \mathcal L_\mathrm{eff}^{(3)} + \dots \,.
\end{align}
In a given UV completion of the SM, this Lagrangian is generated at the new-physics scale $\Lambda\gtrsim m_{Z'}$. It must then be evolved down to the electroweak scale using RG equations, as discussed further in Section~\ref{sec:RGE}. (In some cases an additional evolution below the weak scale may be required. This will not be discussed here.) At the electroweak scale we relate the various fields to the mass eigenstates of the SM particles defined after EWSB. In this step, the $n$-collinear Higgs doublet is written (in unitary gauge) as
\begin{align}
   \Phi_n = \frac{1}{\sqrt{2}}\,W_n^\dagger
    \left( \begin{array}{c} 0 \\ v + h_n \end{array} \right) ,
\end{align}
where the electroweak Wilson line is given by
\begin{align}
   W_n = P \exp\left[ \frac{ig}{2} \int_{-\infty}^0 ds 
    \left( \begin{array}{cc}
    \frac{c_w^2-s_w^2}{c_w}\,\nb\cdot Z_n +2s_w\,\nb\cdot A_n
     &\quad \sqrt{2}\,\nb\cdot W_n^+ \\
    \sqrt{2}\,\nb\cdot W_n^- & - \frac{1}{c_w}\,\nb\cdot Z_n
    \end{array} \right) (s\nb) \right] .
\end{align}
Here $s_w$ and $c_w$ denote the sine and cosine of the electroweak mixing angle. In addition, the various fermion fields must be rotated to the mass basis by diagonalizing the SM Yukawa matrices. In this process the Wilson coefficients of the fermion operators in (\ref{eq:LagLambda2}) and (\ref{eq:LagLambda3}), which are matrices in generation space, are transformed as  
\begin{align}\label{massbasis}
\begin{aligned}
   \bm{C}_{F_L F_L}
   &\to \bm{U}_f^\dagger\,\bm{C}_{F_L F_L} \bm{U}_f \equiv {\bf C}_{f_L f_L} \,, \\
   \bm{C}_{f_R f_R}
   &\to \bm{W}_f^\dagger\,\bm{C}_{f_R f_R} \bm{W}_f \equiv {\bf C}_{f_R f_R} \,, \\
   {\bm{C}}_{F_L f_R}
   &\to \bm{U}_f^\dagger\,{\bm{C}}_{F_L f_R} \bm{W}_f \equiv {\bf C}_{f_L f_R} \,, \\
\end{aligned}
\end{align}
where $f_L$ (with a lower case) now refers to one of the two members of the left-handed doublet $F_L$, and $\bm{U}_f$ and $\bm{W}_f$ with $f=u,d,e$ denote the rotation matrices transforming the left-handed and right handed fermions from the weak to the mass basis. In order not to clutter our notation too much, we use the same symbol but with a straight ``C'' instead of the slanted ``$C$'' for the Wilson coefficients in the mass basis.

\subsection[$Z'$ decays into fermion pairs]{\boldmath $Z'$ decays into fermion pairs}

We begin with the decays into two fermions. The decay amplitude $Z'\to f^i \bar f^j$, where $i,j$ are flavor indices, is given by
\begin{align}\label{eq:fermionFF}
\begin{aligned}
   {\cal M}(Z'\to f_i\bar f_j) 
    = \varepsilon_{Z'}^\mu\,\bar u_{s_1}(p_1)\,\bigg[ 
   & \gamma_\mu^\perp \left( {\rm C}_{f_L f_L}^{\,ij} P_L + {\rm C}_{f_R f_R}^{\,ij} P_R \right) \\
   &\! + \frac{v}{\sqrt2 m_{Z'}}\,\Pi_\mu
    \left( {\rm C}_{f_L f_R}^{\,ij} P_R + {\rm C}_{f_L f_R}^{\,ji\,*} P_L \right)
    + \dots \bigg] P_\nb\,v_{s_2}(p_2) \,,
\end{aligned}
\end{align}
where $s_1$ and $s_2$ denote the spins of the fermions, and the dots indicate higher-order power-suppressed terms. We assume that the fermion $f_i$ moves along the $z$-direction and the anti-fermion $\bar f_j$ moves in the opposite direction. Note that the spinor product of the two highly energetic fermions makes the leading contributions to these amplitudes scale like one power of the hard scale $m_{Z'}$. For the decay rates of an unpolarized $Z'$ boson into fermion pairs with specific helicities, we obtain (with $A,B=L,R$ and $A\ne B$)
\begin{align}\label{eq:fermionrate}
\begin{aligned}
   \Gamma(Z'\to f_A^i\bar f_A^j) 
   &= \frac{N_c^f m_{Z'}}{24\pi}\,
    \lambda^{1/2}\bigg(\frac{m_i^2}{m_{Z'}^2},\frac{m_j^2}{m_{Z'}^2}\bigg) 
    \left| {\rm C}_{f_A f_A}^{\,ij} \right|^2 
    \left[ 1 + {\cal O}\bigg( \frac{v^2}{m_{Z'}^2} \bigg) \right] , \\
   \Gamma(Z'\to f_A^i\bar f_B^j) 
   &= \frac{N_c^f m_{Z'}}{24\pi}\,\frac{v^2}{4m_{Z'}^2}\,
    \lambda^{1/2}\bigg(\frac{m_i^2}{m_{Z'}^2},\frac{m_j^2}{m_{Z'}^2}\bigg) 
    \left| {\rm C}_{f_A f_B}^{\,ij} \right|^2 
    \left[ 1 + {\cal O}\bigg( \frac{v^2}{m_{Z'}^2} \bigg) \right] , 
\end{aligned}
\end{align}
where ${\bf C}_{f_R f_L}\equiv{\bf C}_{f_L f_R}^\dagger$. The overall color factor is $N_c^f=3$ for decays into quarks and $N_c^f=1$ for decays into leptons. Note that the opposite-chirality rates are suppressed, relative to the same-chirality rates, by a factor of order $v^2/m_{Z'}^2$. In each case there are higher-order corrections suppressed by $v^2/m_{Z'}^2$ from higher-order operators in the SCET$_{\rm BSM}$ expansion. These include corrections involving the masses $m_i$ and $m_j$ of the final-state particles. In light of this, it would be consistent to replace the phase-space function $\lambda(x_1,x_2)=(1-x_1-x_2)^2-4x_1 x_2$ in the above expression by~1. 

\subsection[$Z'$ decays into $hV$ final states]{\boldmath $Z'$ decays into $hV$ final states}

Without loss of generality we parameterize the $Z'\to hV$ decay amplitudes (with $V=Z,\gamma$) in terms of Lorentz-invariant form factors characterizing the different polarization states, i.e.\
\begin{align}\label{ZptoVgamma}
\begin{aligned}
   {\cal M}(Z'\to hZ) 
   &= m_{Z'} \bigg[
    \Pi\cdot\varepsilon_{Z'}\,m_Z\,\frac{n\cdot\varepsilon_Z^*}{n\cdot p_Z}\,F_L^{hZ}
    + g_{\mu\nu}^\perp\,\varepsilon_{Z'}^\mu\,\varepsilon_Z^{\nu *} F_T^{\parallel hZ}
    + \epsilon_{\mu\nu}^\perp\,\varepsilon_{Z'}^\mu\,\varepsilon_Z^{\nu *} F_T^{\perp hZ}
    \bigg] \,, \\
   {\cal M}(Z'\to h\gamma) 
   &= m_{Z'} \bigg[
    g_{\mu\nu}^\perp\,\varepsilon_{Z'}^\mu\,\varepsilon_\gamma^{\nu *} F_T^{\parallel h\gamma}
    + \epsilon_{\mu\nu}^\perp\,\varepsilon_{Z'}^\mu\,\varepsilon_\gamma^{\nu *} 
    F_T^{\perp h\gamma} \bigg] \,, 
\end{aligned}
\end{align}
where we always choose the reference frame such that the first (second) particle in the final state moves along the positive (negative) $z$ axis, aligned with the reference vectors $n$ ($\nb$). At leading order in $\lambda=v/m_{Z'}$ the form factors are given by 
\begin{align}
\begin{aligned}
   F_L^{hZ} &= C_{\phi\phi} + {\cal O}\bigg(\frac{v^2}{m_{Z'}^2}\bigg) \,, \\
   F_T^{\sigma hZ} 
   &= - \frac{m_Z}{m_{Z'}}\,\Re e \left( c_w^2\,C_{W\phi}^\sigma + s_w^2\,C_{B\phi}^\sigma \right) 
    + {\cal O}\bigg(\frac{v^3}{m_{Z'}^3}\bigg) \,; \quad \sigma =\,\, \parallel,\perp  
\end{aligned}
\end{align}
and
\begin{align}
   F_T^{\sigma h\gamma} = - \frac{m_Z}{m_{Z'}}\,s_w\spac c_w\,
    \Re e \left( C_{W\phi}^\sigma - C_{B\phi}^\sigma \right) 
    + {\cal O}\bigg(\frac{v^3}{m_{Z'}^3}\bigg) \,; \quad \sigma =\,\, \parallel,\perp .
\end{align}
Note that the transverse form factors are suppressed by a factor $m_Z/m_{Z'}$, reflecting the fact that they descend from operators of subleading power in the SCET$_{\rm BSM}$ expansion. 

For the unpolarized $Z'\to Vh$ decay rates we obtain
\begin{align}\label{eq47}
\begin{aligned}
   \Gamma(Z'\to hZ) 
   &= \frac{m_{Z'}}{48\pi}\,\lambda^{1/2}\bigg(\frac{m_Z^2}{m_{Z'}^2},\frac{m_h^2}{m_{Z'}^2}\bigg) 
    \left\{ \big| F_L^{hZ} \big|^2 + 2 \left[ \big| F_T^{\parallel hZ} \big|^2 
    + \big| F_T^{\perp hZ} \big|^2 \right] \right\} , \\
   \Gamma(Z'\to h\gamma) 
   &= \frac{m_{Z'}}{24\pi} \left( 1 - \frac{m_h^2}{m_{Z'}^2} \right) 
    \left[ \big| F_T^{\parallel h\gamma} \big|^2 + \big| F_T^{\perp h\gamma} \big|^2 \right] .
\end{aligned}
\end{align}
The factor 2 in the first result accounts for the two transverse polarization states of the final-state vector meson. Note that the form factors can be complex functions in higher orders of perturbation theory, and we therefore use absolute value signs in the expression for the rates.

\subsection[$Z'$ decays into two gauge bosons]{\boldmath $Z'$ decays into two gauge bosons}

We finally turn to the decays of a $Z'$ boson into a pair of gauge bosons. The corresponding decay amplitudes can be expressed in terms of the form-factor decomposition
\begin{align}\label{eq46}
\begin{aligned}
   {\cal M}(Z'\to V_1 V_2) 
   &= m_{Z'} \bigg[ \Pi\cdot\varepsilon_{Z'}\,
    m_{V_1}\,\frac{\nb\cdot\varepsilon_{V_1}^*}{\nb\cdot p_{V_1}}\,
    m_{V_2}\,\frac{n\cdot\varepsilon_{V_2}^*}{n\cdot p_{V_2}}\,F_{LL}^{V_1 V_2} \\
   &\hspace{1.4cm} + m_{V_1}\,\frac{\nb\cdot\varepsilon_{V_1}^*}{\nb\cdot p_{V_1}}\,
    \varepsilon_{Z'}^\mu\,\varepsilon_{V_2}^{*\nu} 
    \left( g_{\mu\nu}^\perp\,F_{LT}^{\parallel V_1 V_2} 
    + \epsilon_{\mu\nu}^\perp\,F_{LT}^{\perp V_1 V_2} \right) \\
   &\hspace{1.4cm} + m_{V_2}\,\frac{n\cdot\varepsilon_{V_2}^*}{n\cdot p_{V_2}}\,
    \varepsilon_{Z'}^\mu\,\varepsilon_{V_1}^{*\nu} 
    \left( g_{\mu\nu}^\perp\,F_{TL}^{\parallel V_1 V_2} 
    - \epsilon_{\mu\nu}^\perp\,F_{TL}^{\perp V_1 V_2} \right) \!\bigg] \,,
\end{aligned}
\end{align}
where we assume that particle $V_1$ moves along the positive $z$ axis. The indices $L$ and $T$ on the form factors denote longitudinal and transverse polarizations of the two vector bosons $V_1$ and $V_2$, respectively. As a consequence of the Landau-Yang theorem \cite{Landau:1948kw,Yang:1950rg}, at least one of the final-state vector bosons needs to be longitudinally polarized. (This will no longer be true if the heavy vector meson carries color, see Section~\ref{sec:nonsinglet}.)

We now list our results for the various form factors for the three possible final states $W^+ W^-$, $ZZ$ and $Z\gamma$, obtained from the effective Lagrangian up to subleading order in $\lambda$. The purely longitudinal polarization configuration is only allowed for the decay $Z'\to W^+ W^-$. We find
\begin{align}
   F_{LL}^{W^+W^-} = - C_{\phi\phi} + {\cal O}\bigg( \frac{v^2}{m_{Z'}^2} \bigg) \,, \qquad
   F_{LL}^{ZZ} = F_{LL}^{Z\gamma} = 0 \,.
\end{align}
An on-shell photon does not have a longitudinal polarization state, and for the $ZZ$ final state the form factor $F_{LL}^{ZZ}$ vanishes due to Bose symmetry. The contributions to the decay amplitude corresponding to mixed polarization states of the final-state particles arise at subleading order in power counting. We find
\begin{align}
\begin{aligned}
   F_{LT}^{\sigma W^+W^-} 
   &= \frac{m_W}{m_{Z'}}\,C_{W\phi}^\sigma + {\cal O}\bigg( \frac{v^3}{m_{Z'}^3} \bigg) \,, \\
   F_{LT}^{\sigma ZZ} 
   &= i\,\frac{m_Z}{m_{Z'}}\,\Im m \left( c_w^2\,C_{W\phi}^\sigma + s_w^2\,C_{B\phi}^\sigma \right) 
    + {\cal O}\bigg( \frac{v^3}{m_{Z'}^3} \bigg) \,, \\
   F_{LT}^{\sigma Z\gamma} 
   &= i\,\frac{m_Z}{m_{Z'}}\,s_w\spac c_w\,
    \Im m \left( C_{W\phi}^\sigma - C_{B\phi}^\sigma \right) 
    + {\cal O}\bigg( \frac{v^3}{m_{Z'}^3} \bigg) \,,
\end{aligned}
\end{align}
as well as 
\begin{align}\label{eq3.14}
   F_{TL}^{\sigma W^+W^-} = - \left( F_{LT}^{\sigma W^+W^-} \right)^* , \qquad
   F_{TL}^{\sigma ZZ} = F_{LT}^{\sigma ZZ} \,, \qquad
   F_{TL}^{\sigma Z\gamma} = 0 \,.
\end{align}
As previously $\sigma=\,\,\parallel,\perp$. Note that the $Z'\to ZZ$ and $Z'\to Z\gamma$ form factors are purely imaginary and therefore CP odd. By searching for these decay modes, is it possible to probe for the presence of CP-violating interactions in the UV theory.

It is instructive to compare our result for the $Z'\to ZZ$ decay amplitude with the findings of \cite{Keung:2008ve}, where the authors parameterized the couplings of the two $Z$ bosons to the $Z'$ resonance in terms of the two operators 
\begin{align}
   {\cal L}_{\rm eff} = f_4\,Z_\mu'(\partial_\nu Z^\mu) Z^\nu
    + f_5\,\epsilon^{\mu\nu\rho\sigma} Z'_\mu Z_\nu (\partial_\rho Z_\sigma) 
\end{align}
in the broken phase of the electroweak symmetry. For the corresponding contributions to the form factors we find
\begin{align}
   F^{\parallel ZZ}_{LT} = \frac{i f_4 m_{Z'}}{2 m_Z} \,, \qquad
   F^{\perp ZZ}_{LT} = \frac{i f_5 m_{Z'}}{2m_Z} \,.
\end{align}
It follows that the coefficients $f_4$ and $f_5$ correspond to linear combinations of the imaginary parts of our Wilson coefficients $C_{W\phi}^\sigma$ and $C_{B\phi}^\sigma$. 

In terms of the form factors defined in (\ref{eq46}), the $Z'\to V_1 V_2$ decay rates of an unpolarized $Z'$ boson are obtained as
\begin{align}
   \Gamma(Z'\to W^+W^-) 
   &= \frac{m_{Z'}}{48\pi}\,\sqrt{ 1 - \frac{4m_W^2}{m_{Z'}^2}}
    \left\{ \big| F_{LL}^{W^+W^-} \big|^2 + 4 \left[ \big| F_{LT}^{\parallel W^+W^-} \big|^2 
    + \big| F_{LT}^{\perp W^+W^-} \big|^2 \right] \right\} , \notag\\[-1mm]
   \Gamma(Z'\to ZZ) 
   &= \frac{m_{Z'}}{24\pi}\,\sqrt{ 1 - \frac{4m_Z^2}{m_{Z'}^2}}
    \left[ \big| F_{LT}^{\parallel ZZ} \big|^2 + \big| F_{LT}^{\perp ZZ} \big|^2 \right] , \\[0.5mm]
   \Gamma(Z'\to Z\gamma) 
   &= \frac{m_{Z'}}{24\pi} \left( 1 - \frac{m_Z^2}{m_{Z'}^2} \right)
    \left[ \big| F_{LT}^{\parallel Z\gamma} \big|^2 + \big| F_{LT}^{\perp Z\gamma} \big|^2 \right] , \notag
\end{align}
where we have used (\ref{eq3.14}) included a symmetry factor of 1/2 in the second case.

\renewcommand{\theequation}{4.\arabic{equation}}
\setcounter{equation}{0}

\section{Resummation of large logarithms}
\label{sec:RGE}

A key strength of any EFT framework is that it allows for a systematic resummation of large logarithms present in multi-scale problems, which could otherwise spoil the convergence of the perturbative expansion. This is achieved by solving the RG evolution equations of the theory. We now discuss the resummation of the large (single and double) logarithms of the ratio $m_{Z'}^2/m_{\rm SM}^2$ for two representative examples, focusing for simplicity on the operators arising in the leading-power effective Lagrangian (\ref{eq:LagLambda2}). A detailed discussion of the derivation of the anomalous dimensions governing the scale dependence of the Wilson coefficients in the effective Lagrangian, both at leading and subleading order in $\lambda$, has been presented in Ref.~\cite{Alte:2018nbn}.

The Wilson coefficients $C_{\phi\phi}$ and $\bm{C}_{\psi\psi}$ (considered as a matrix in generation space) obey the RG evolution equations 
\begin{align}\label{RGEslam2}
\begin{aligned}
   \mu\,\frac{d}{d\mu}\,C_{\phi\phi}(m_{Z'},\mu) 
   &= \Gamma_{\phi\phi}(\mu)\,C_{\phi\phi}(m_{Z'},\mu) \,, \\
   \mu\,\frac{d}{d\mu}\,\bm{C}_{\psi\psi}(m_{Z'},\mu) 
   &= \bm{\Gamma}_{\psi\psi}(\mu)\otimes\bm{C}_{\psi\psi}(m_{Z'},\mu) \,,
\end{aligned}
\end{align}
where the $\otimes$ symbol in the second equation signals that a proper ordering of the two matrices must be taken into account. To all orders in perturbation theory the anomalous dimensions take the form \cite{Becher:2009cu,Becher:2009qa}
\begin{align}\label{gammares}
   \Gamma_{\phi\phi}(\mu) 
   &= \left( \frac14\,\gamma_\mathrm{cusp}^{(1)} + \frac34\,\gamma_\mathrm{cusp}^{(2)} \right)
    \left( \ln\frac{m_{Z'}^2}{\mu^2} - i\pi \right) + 2\gamma^\phi \,, \notag\\
   \bm{\Gamma}_{Q_L Q_L}(\mu) 
   &= \left( \frac{1}{36}\,\gamma_\mathrm{cusp}^{(1)} + \frac34\,\gamma_\mathrm{cusp}^{(2)}
    + \frac43\,\gamma_\mathrm{cusp}^{(3)} \right) 
    \left( \ln\frac{m_{Z'}^2}{\mu^2} - i\pi \right) + \{ \bm{\gamma}^{Q_L}, \,.\, \} \,, \notag\\
   \bm{\Gamma}_{L_L L_L}(\mu) 
   &= \left( \frac14\,\gamma_\mathrm{cusp}^{(1)} + \frac34\,\gamma_\mathrm{cusp}^{(2)} \right) 
    \left( \ln\frac{m_{Z'}^2}{\mu^2} - i\pi \right) + \{ \bm{\gamma}^{L_L}, \,.\, \} \,, \\
   \bm{\Gamma}_{q_R q_R}(\mu) 
   &= \left( e_q^2\,\gamma_\mathrm{cusp}^{(1)} + \frac43\,\gamma_\mathrm{cusp}^{(3)} \right) 
    \left( \ln\frac{m_{Z'}^2}{\mu^2} - i\pi \right) + \{ \bm{\gamma}^{q_R}, \,.\, \} \,; \quad q=u,d , 
    \notag\\
   \bm{\Gamma}_{e_R e_R}(\mu) 
   &= \gamma_\mathrm{cusp}^{(1)} \left( \ln\frac{m_{Z'}^2}{\mu^2} - i\pi \right) 
    + \{ \bm{\gamma}^{e_R}, \,.\, \} \,. \notag
\end{align}
Terms without a boldface symbol are proportional to the unit matrix in generation space. The notation $\{\bm{\gamma}^F, \,.\,\}$ means that 
\begin{align}
   \{ \bm{\gamma}^F, \,.\, \}\otimes\bm{C}_{\psi\psi}
   \equiv \bm{\gamma}^F \bm{C}_{\psi\psi} + \bm{C}_{\psi\psi}\,\bm{\gamma}^F \,.
\end{align}
In general, the cusp anomalous dimensions $\gamma_{\rm cusp}^{(r)}$ and the single-particle anomalous dimensions $\gamma^i$ depend on the three gauge couplings $\alpha_1=g^{\prime\,2}/(4\pi)$, $\alpha_2=g^2/(4\pi)$ and $\alpha_3=\alpha_s$, the quartic scalar coupling $\lambda_\phi$ and the Yukawa couplings. Up to two-loop order, however, the cusp anomalous dimension for the gauge group $G_r$ only depends on the corresponding coupling $\alpha_r$ \cite{Korchemsky:1987wg,Korchemskaya:1992je,Jantzen:2005az}. Note, in particular, that there are no contributions to the cusp anomalous dimensions from Yukawa interactions, because the relevant vertex graphs are found to be power suppressed. Explicitly, one finds
\begin{align}\label{gammacusp}
\begin{aligned}
   \gamma_{\rm cusp}^{(1)} 
   &= \frac{\alpha_1}{\pi} - \frac{17}{6} \left( \frac{\alpha_1}{\pi} \right)^2 + \dots \,, \\ 
   \gamma_{\rm cusp}^{(2)}
   &= \frac{\alpha_2}{\pi} + \left( 2 - \frac{\pi^2}{6} \right) 
    \left( \frac{\alpha_2}{\pi} \right)^2 + \dots \,, \\ 
   \gamma_{\rm cusp}^{(3)} 
   &= \frac{\alpha_3}{\pi} + \left( \frac{47}{12} - \frac{\pi^2}{4} \right) 
    \left( \frac{\alpha_3}{\pi} \right)^2 + \dots \,.
\end{aligned}   
\end{align}
To one-loop order, the relevant single-particle anomalous dimensions read \cite{Alte:2018nbn}
\begin{align}\label{singleparticle}
   \gamma^\phi 
   &= - \frac{\alpha_1}{4\pi} - \frac{3\alpha_2}{4\pi} 
    + \sum_f\,\frac{N_c^f y_f^2}{16\pi^2} + \dots \,, \notag\\[-2mm]
   \bm{\gamma}^{Q_L} 
   &= - \frac{\alpha_1}{144\pi} - \frac{9\alpha_2}{16\pi} - \frac{\alpha_3}{\pi} 
    + \frac{1}{32\pi^2} \left( \bm{y}_u\spac\bm{y}_u^\dagger 
    + \bm{y}_d\spac\bm{y}_d^\dagger \right) + \dots , \notag\\
   \bm{\gamma}^{L_L} 
   &= - \frac{\alpha_1}{16\pi} - \frac{9\alpha_2}{16\pi} 
    + \frac{1}{32\pi^2}\,\bm{y}_e\spac\bm{y}_e^\dagger + \dots \,, \\
   \bm{\gamma}^{q_R} 
   &= - e_q^2\,\frac{\alpha_1}{4\pi} - \frac{\alpha_3}{\pi} 
    + \frac{1}{16\pi^2}\,\bm{y}_q^\dagger\spac\bm{y}_q + \dots \,; \quad q=u,d , \notag\\
   \bm{\gamma}^{e_R} 
   &= - \frac{\alpha_1}{4\pi}  
    + \frac{1}{16\pi^2}\,\bm{y}_e^\dagger\spac\bm{y}_e + \dots \,, \notag
\end{align}
where in the first expression the sum runs over the different fermion species $f$ (the six quark flavors $u,d,s,c,b,t$ and the three charged leptons $e,\mu,\tau$), $\bm{y}_i$ with $i=u,d,e$ are the SM Yukawa matrices, and $y_f$ denotes the Yukawa coupling of the fermion $f$ in the mass basis. When the Wilson coefficients $\bm{C}_{\psi\psi}$ are transformed into the mass basis, as shown in (\ref{massbasis}), the Yukawa matrices in (\ref{singleparticle}) are brought to diagonal form with the exception of $\bm{\gamma}^{Q_L}$, for which one of the two terms in $(\bm{y}_u\spac\bm{y}_u^\dagger+\bm{y}_d\spac\bm{y}_d^\dagger)$ still contains off-diagonal entries. For the coefficients ${\bf C}_{u_L}$ and ${\bf C}_{d_L}$ one finds that this quantity is transformed into
\begin{align}
\begin{aligned}
   & {\bf C}_{u_L}: \quad \text{diag}(y_u^2,y_c^2,y_t^2)
    + \bm{V}\,\text{diag}(y_d^2,y_s^2,y_b^2)\,\bm{V}^\dagger \,, \\
   & {\bf C}_{d_L}: \quad \bm{V}^\dagger\,\text{diag}(y_u^2,y_c^2,y_t^2)\,\bm{V}
    + \text{diag}(y_d^2,y_s^2,y_b^2) \,,
\end{aligned}   
\end{align}
where $\bm{V}=\bm{U}_u^\dagger\spac\bm{U}_d$ is the CKM matrix. In practice, it is an excellent approximation to neglect all eigenvalues of the Yukawa matrices except for the top-quark Yukawa $y_t\approx 1$. One then obtains in the mass basis 
\begin{align}\label{eq61}
\begin{aligned}
   \gamma^\phi 
   &= - \frac{\alpha_1}{4\pi} - \frac{3\alpha_2}{4\pi} + \frac{3 y_t^2}{16\pi^2} + \dots \,, \\[-2mm]
   \bm{\gamma}^{u_L} 
   &= - \frac{\alpha_1}{144\pi} - \frac{9\alpha_2}{16\pi} - \frac{\alpha_3}{\pi} 
    + \frac{y_t^2}{32\pi^2}\,\text{diag}(0,0,1) + \dots \,, \\
   \bm{\gamma}^{d_L} 
   &= - \frac{\alpha_1}{144\pi} - \frac{9\alpha_2}{16\pi} - \frac{\alpha_3}{\pi} 
    + \frac{y_t^2}{32\pi^2}\,\big( V_{.\,3}^\dagger\spac V_{3\,.} \big) + \dots \,, \\
   \bm{\gamma}^{u_R} 
   &= - \frac{\alpha_1}{9\pi} - \frac{\alpha_3}{\pi} 
    + \frac{y_t^2}{16\pi^2}\,\text{diag}(0,0,1) + \dots \,,
\end{aligned}   
\end{align}
whereas the Yukawa contributions to $\bm{\gamma}^{L_L}$, $\bm{\gamma}^{d_R}$ and $\bm{\gamma}^{e_R}$ can be dropped. In the third relation $(V_{.\,3}^\dagger\spac V_{3\,.})$ denotes the matrix with elements $(V_{.\,3}^\dagger\spac V_{3\,.})_{ij}\equiv V_{ti}^*\spac V_{tj}$.

We do not present here a detailed discussion of the anomalous dimensions of the ${\cal O}(\lambda^3)$ operators defined in (\ref{lam3ops}), which is complicated by the fact that these operators contain a zero-momentum Higgs field. Focussing on QCD evolution only, we find that the Wilson coefficients ${\bf C}_{q_L q_R}$ of the subleading fermionic operators containing quark fields obey an evolution equation analogous to (\ref{RGEslam2}) with the anomalous dimension \cite{Becher:2009cu,Becher:2009qa}
\begin{align}
   \Gamma_{q_L q_R}(\mu) 
   = \frac43\,\gamma_\mathrm{cusp}^{(3)} \left( \ln\frac{m_{Z'}^2}{\mu^2} - i\pi \right) - \frac{2\alpha_3}{\pi} \,,
\end{align}
which is identical to the QCD part of the anomalous dimensions $\bm{\Gamma}_{Q_L Q_L}$ and $\bm{\Gamma}_{q_R q_R}$ in (\ref{gammares}).
 
In order to illustrate the typical size of resummation effects, we solve the RG evolution equations numerically for two representative examples: the decays $Z'\to t\bar t$ and $Z'\to W^+ W^-$, both evaluated at leading order in the expansion in powers of $\lambda$ and for a mass $m_{Z'}=3$\,TeV of the new $Z'$ boson. A consistent solution at leading order in RG-improved perturbation theory requires that one uses the two-loop expressions for the cusp anomalous dimensions in (\ref{gammacusp}) and for the  $\beta$-functions of the SM gauge couplings \cite{Mihaila:2012pz}, while the one-loop expressions are sufficient for all other anomalous dimensions 
and for the $\beta$-function of the top-quark Yukawa coupling. The latter one is given by \cite{Grzadkowski:1987tf}
\begin{align}
   \mu\,\frac{d}{d\mu}\,y_t = \frac{9\spac y_t^3}{32\pi^2}
    - y_t \left( \frac{17\alpha_1}{48\pi} + \frac{9\alpha_2}{16\pi} + \frac{2\alpha_3}{\pi} \right) .
\end{align}
For the case of $Z'\to t\bar t$ the relevant Wilson coefficients are ${\rm C}_{u_L u_L}^{33}(m_{Z'},\mu_t)$ and ${\rm C}_{u_R u_R}^{33}(m_{Z'},\mu_t)$, where $\mu_t\approx m_t$ is a characteristic low-energy scale of the decay process. From (\ref{eq61}) it follows that both coefficients obey a diagonal RG equation (in the approximation where we neglect all Yukawa couplings other than $y_t$). We define
\begin{align}
    C_X(m_{Z'},\mu_l) = C_X(m_{Z'},\mu_h)\,U_X(\mu_h,\mu_l)
\end{align}
for any Wilson coefficient at a low scale $\mu_l$, where the evolution factor $U_X(\mu_h,\mu_l)$ encodes the scale evolution from the new-physics scale $\mu_h\sim m_{Z'}$ down to a characteristic low-energy scale $\mu_l$. From a numerical solution of the RG equations we find
\begin{align}
   U_{u_L u_L}^{33}(m_{Z'},m_t) \approx 0.78\,e^{0.45\spac i} \,, \qquad
   U_{u_R u_R}^{33}(m_{Z'},m_t) \approx 0.79\,e^{0.39\spac i} \,.
\end{align}
Depending on which Wilson coefficient dominates, the $Z'\to t\bar t$ decay rate drops by a factor of $|U_{q_L q_L}^{33}|^2\approx 0.61$ or $|U_{u_R u_R}^{33}|^2\approx 0.63$, which is a sizable correction, largely driven by QCD. 

\begin{figure}
\centering
\includegraphics[scale=0.7]{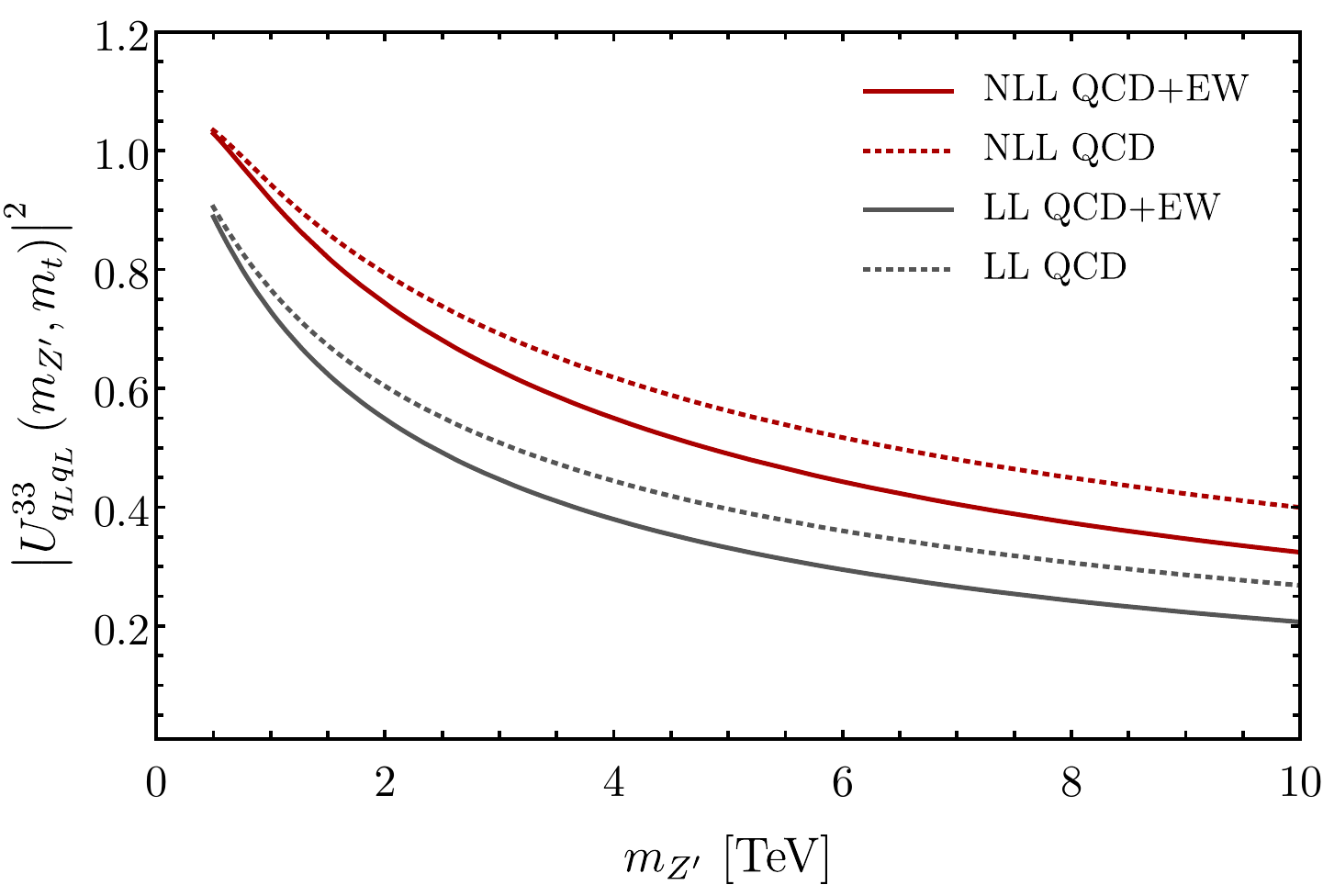}
\caption{Resummation effects for the Wilson coefficient $|{\rm C}_{u_L u_L}^{33}|^2$ using different approximations: The red curves show the effects obtained from solving the RG equations with all terms included, while the gray curves are obtained by keeping the cusp terms only. In both cases the solid curves include electroweak evolution effects and the contributions involving the top-quark Yukawa coupling, whereas the dotted curves correspond to QCD evolution only.}
\label{fig:RGE}
\end{figure}

It is interesting to compare these results to the ones obtained from parton showers. We simulate the effect of a typical QCD parton shower by including only the terms proportional to $\gamma_{\rm cusp}^{(3)}$ in the anomalous dimensions of the Wilson coefficients. In this approximation the evolution factors turn out to be
\begin{align}
   U_{u_L u_L}^{33}(m_{Z'},m_t) \big|_{\gamma_{\rm cusp}^{(3)}} 
   = U_{u_R u_R}^{33}(m_{Z'},m_t) \big|_{\gamma_{\rm cusp}^{(3)}}
   \approx 0.70\,e^{0.36\spac i}\,,
\end{align}
and the rate thus drops by a factor of approximately 0.49. The difference can be attributed to the effects of subleading (single) logarithms as well electroweak evolution effects, where the two contributions partially compensate each other, as we demonstrate in Figure~\ref{fig:RGE}. There we show the complete RG evolution effects (red curves) and compare them to the solution where only the cusp terms in the anomalous dimensions are kept (gray curves). We see that including the subleading logarithms attenuates the running while including electroweak effects amplifies it. It should also be noted that the difference between the curves are more or less constant, making the relative difference between the effects more pronounced with growing scale separation.

While in the previous example QCD evolution effects are the dominant ones, there are other observables for which the evolution driven by electroweak and Yukawa interactions can be similarly important. The $Z'\to W^+W^-$ decay rate is, at leading power in $\lambda$, determined by the Wilson coefficient $C_{\phi\phi}(m_{Z'},\mu)$. Evolving this coefficient from the new-physics scale down to $\mu_l=m_W$, we obtain
\begin{align}
   U_{\phi\phi}(m_{Z'},m_W) \approx 0.85\,e^{0.10\spac i} \,.
\end{align}
The resummation effects reduce the decay rate by a factor of $|U_{\phi\phi}|^2\approx 0.72$. This effect is sizable, even without QCD evolution, because of the large top-quark Yukawa coupling contributing to the anomalous dimension $\Gamma_{\phi\phi}$ entering through the quantity $\gamma^\phi$ in (\ref{gammares}).

\renewcommand{\theequation}{5.\arabic{equation}}
\setcounter{equation}{0}

\section{Extension to non-singlet resonances}
\label{sec:nonsinglet}

The theoretical framework described in the previous sections can be straightforwardly extended to the case of heavy vector resonances charged under the SM gauge group. We now briefly outline the modifications in the construction of the effective SCET$_{\rm BSM}$ Lagrangian for the example of a Kaluza-Klein gluon arising in extra-dimensional extensions of the SM, i.e., for a heavy color-octet vector resonance $G'$.

\subsection{Operator basis}

As before the leading-order operators arise at ${\cal O}(\lambda^2)$. They are constructed in analogy to the treatment discussed in Section~\ref{subsec:2.3}. Color conservation requires that couplings to fermions exist only in the quark sector. The relevant operators are of the form 
\begin{align}
   O_{\psi\psi}^{ij} 
   = G_{v\mu}^{\prime a} \big( \bar\Psi^i_n\gamma_\perp^\mu\spac t^a\spac\Psi^j_\nb
    + \bar\Psi^i_\nb\gamma_\perp^\mu\spac t^a\spac\Psi^j_n \big) \,, 
\end{align}
where $t^a$ are the generators of $SU(3)_c$, and the fermion fields can be left-handed or right-handed quark fields. There is no operator analogous to $O_{\phi\phi}$ in (\ref{Ophiphi}), because the Higgs field does not carry a color index. On the other hand, it is possible to construct non-vanishing operators of the form (\ref{ZpAA}), in which the heavy color-octet resonance couples to a pair of gauge bosons. They are (with $b\equiv 1$)
\begin{align}
\begin{aligned}
   \tilde O_{GB}^\parallel &= i m_{G'}\,\Pi\cdot G_v^{\prime a}\,g_{\mu\nu}^\perp\,
    \big( \G_n^{\perp\mu,a}\,\B_\nb^{\perp\nu,b} - \G_\nb^{\perp\mu,a}\,\B_n^{\perp\nu,b} \big) \,, \\
   \tilde O_{GB}^\perp &= i m_{G'}\,\Pi\cdot G_v^{\prime a}\,\epsilon_{\mu\nu}^\perp\,
    \big( \G_n^{\perp\mu,a}\,\B_\nb^{\perp\nu,b} + \G_\nb^{\perp\mu,a}\,\B_n^{\perp\nu,b} \big) \,,
\end{aligned}
\end{align}
where the effective fields $\G$ and $\B$ represent a gluon and a hypercharge gauge boson, respectively. Because these fields refer to different particles, it is possible to construct boson bilinears that are odd under the exchange of $n$ and $\nb$, as required by the symmetry properties of $\Pi^\mu$ in (\ref{Pidef}). Note that these hermitian operators mediate CP-odd interactions, as indicated by the tilde symbol. This exhausts all options at $\mathcal{O}(\lambda^2)$. The most general effective Lagrangian at this order is therefore (a sum over repeated indices in implied)
\begin{align}
   \mathcal{L}_\mathrm{eff}^{(2)}
   = \sum_\psi C_{\psi\psi}^{ij}(m_{G'},\mu)\,O^{ij}_{\psi\psi}(\mu) 
    + \sum_{\sigma=\parallel,\perp} \tilde C_{GB}^\sigma(m_{G'},\mu)\,\tilde O_{GB}^\sigma(\mu) \,,
\end{align}
where the sum in the first term runs over all quark multiplets of the SM, and the Wilson coefficients $C_{\psi\psi}^{ij}$ form the entries of $3\times 3$ hermitian matrices. The coefficients $C_{GB}^\sigma$, on the other hand, are real quantities. This effective Lagrangian mediates the two-jet decays $G'\to q\bar q$, $G'\to g\gamma$ and $G'\to g Z$, where the $Z$ boson carries transverse polarization. The latter two transitions are CP odd and can be used to probe for CP-violating interactions in the UV completion of the SM.

At subleading order in power counting there exists again a long list of possible operators, but at lowest order in perturbation theory we only need those constructed in analogy with those shown in (\ref{lam3ops}). We find that the relevant operators are 
\begin{align}
   O_{Q_L q_R}^{ij} 
   &= \frac{\Pi\cdot G^{\prime a}}{m_{G'}}\,\big( \bar Q_{L,n}^i\Phi_0\spac t^a q_{R,\nb}^j 
    - \bar Q_{L,\nb}^i\Phi_0\spac t^a q_{R,n}^j \big) \,, \notag\\[-1mm]
   O_{G\phi}^\parallel &= g_{\mu\nu}^\perp\,G_v^{\prime\mu,a}\,\big( 
    \Phi_n^\dagger\Phi_0\spac\G_\nb^{\perp\nu,a} 
    + \Phi_\nb^\dagger\Phi_0\spac\G_n^{\perp\nu,a} \big) \,, \\[1mm]
   O_{G\phi}^\perp &= \epsilon_{\mu\nu}^\perp\,G_v^{\prime\mu,a}\,\big( 
    \Phi_n^\dagger\Phi_0\spac\G_\nb^{\perp\nu,a} 
    - \Phi_\nb^\dagger\Phi_0\spac\G_n^{\perp\nu,a} \big) \,. \notag
\end{align}
We write the effective Lagrangian at order ${\cal O}(\lambda^3)$ as 
\begin{align}
   \mathcal L_\mathrm{eff}^{(3)}
   = \sum_{q=u,d} C_{Q_L q_R}^{ij}(m_{G'},\mu)\,O_{Q_L q_R}^{ij}(\mu) 
    + \sum_{\sigma=\parallel,\perp} C_{G\phi}^\sigma(m_{G'},\mu)\,O_{G\phi}^\sigma(\mu) 
    + \text{h.c.} + \dots \,,
\end{align}
where the dots refer to the many other operators, which do not contribute at tree level to the amplitudes we consider. For the case where $q_R=u_R$ the replacement $\Phi\to\tilde\Phi$ must be made to ensure gauge invariance. The Wilson coefficients in this Lagrangian are arbitrary complex quantities. The latter two operators mediate the two-jet decays $G'\to g h$ and $G'\to g Z$ with a longitudinally polarized $Z$ boson.

\subsection{Decay amplitudes and rates}

In analogy with the discussion in Section~\ref{sec:matrixelem}, we now briefly discuss the relevant two-body decay amplitudes and the corresponding decay rates. The amplitudes for the decays $G'\to 
q^i\spac\bar q^j$ have the same form as shown in (\ref{eq:fermionFF}), but with a color generator $t^a$ inserted between the two spinors. The decay rates into quark pairs with various chiralities are thus given by the expressions in (\ref{eq:fermionrate}) with $m_{Z'}$ replaced by $m_{G'}$ and the color factor $N_c^f$ replaced by $T_F=\frac12$. At leading order in power counting the resonance $G'$ decays into quarks with equal chiralities.

The decay $G'\to hg$ arises first at subleading order in power counting. The decay amplitude can be parameterized in the same way as the $Z'\to h\gamma$ decay amplitude in (\ref{ZptoVgamma}). For the form factors we obtain
\begin{align}
   F_T^{\sigma hg} = \frac{g_s v}{m_{G'}}\,\Re e\,C_{G\phi}^\sigma 
    + {\cal O}\bigg(\frac{v^3}{m_{Z'}^3}\bigg) \,; \quad \sigma =\,\, \parallel,\perp .
\end{align}
The corresponding decay rate is given by an expression analogous to the second relation in (\ref{eq47}), but with $m_{Z'}$ replaced by $m_{G'}$.

The decays of the resonance $G'$ into two gauge bosons are particularly interesting, because contrary to the decay $Z'\to Z\gamma$ the modes $G'\to g\gamma$ and $G'\to gZ$ can be mediated by operators arising at leading order in power counting. Instead of (\ref{eq46}), we parameterize the corresponding decay amplitudes in the form (with $V=\gamma, Z$)
\begin{align}
\begin{aligned}
   {\cal M}(G'\to gV) 
   &= m_{G'} \bigg[ \Pi\cdot\varepsilon_{G'}\,\varepsilon_g^{*\mu}\,\varepsilon_V^{*\nu}
    \left( g_{\mu\nu}^\perp\,F_{TT}^{\parallel gV} 
    + \epsilon_{\mu\nu}^\perp\,F_{TT}^{\perp gV} \right) \\
   &\hspace{1.4cm} + m_V\,\frac{n\cdot\varepsilon_V^*}{n\cdot p_V}\,
    \varepsilon_{G'}^\mu\,\varepsilon_g^{*\nu} 
    \left( g_{\mu\nu}^\perp\,F_{TL}^{\parallel gV} 
    - \epsilon_{\mu\nu}^\perp\,F_{TL}^{\perp gV} \right) \!\bigg] \,,
\end{aligned}
\end{align}
where the terms in the second line exist only if $V$ is a $Z$ boson. For the form factors we find the expressions
\begin{align}
   F_{TT}^{\sigma g\gamma} = i\spac g_s\spac e\,C_{GB}^\sigma \,, \qquad 
   F_{TT}^{\sigma gZ} = - i\,\frac{s_w}{c_w}\,g_s\spac e\,C_{GB}^\sigma \,,
\end{align}
and
\begin{align}
   F_{TL}^{\sigma gZ} = - i\,\frac{g_s v}{m_{G'}}\,\Im m\,C_{G\phi}^\sigma \,, \qquad 
   F_{TL}^{\sigma g\gamma} = 0 \,.
\end{align}
The fact that these form factors are purely imaginary indicates the fact that the corresponding decay amplitudes are CP odd. For the decay rates we obtain
\begin{align}
\begin{aligned}
   \Gamma(G'\to g\gamma) 
   &= \frac{m_{G'}}{24\pi} 
    \left[ \big| F_{TT}^{\parallel g\gamma} \big|^2 + \big| F_{TT}^{\perp g\gamma} \big|^2 \right] , \\
   \Gamma(G'\to gZ) 
   &= \frac{m_{G'}}{24\pi} \left( 1 - \frac{m_Z^2}{m_{G'}^2} \right)
    \left[ \big| F_{TT}^{\parallel gZ} \big|^2 + \big| F_{TT}^{\perp gZ} \big|^2
    + \big| F_{TL}^{\parallel gZ} \big|^2 + \big| F_{TL}^{\perp gZ} \big|^2 \right] .
\end{aligned}
\end{align}

\subsection{Resummation of large logarithms}

RG resummation effects are more interesting in the case where the decaying heavy resonance is charged under the SM gauge group. We now discuss this for the case of the operators appearing at leading order in power counting. Their anomalous dimensions can be derived from a two-loop master formula for the anomalous dimensions of scattering amplitudes containing both massless and massive partons derived in Refs.~\cite{Becher:2009kw,Becher:2019avh,Ferroglia:2009ep,Ferroglia:2009ii}. We find that
\begin{align}
   \bm{\Gamma}_{Q_L Q_L}(\mu) 
   &= \left( \frac{1}{36}\,\gamma_{\rm cusp}^{(1)} + \frac34\,\gamma_{\rm cusp}^{(2)} 
    - \frac16\,\gamma_{\rm cusp}^{(3)} \right) \left( \ln\frac{m_G^2}{\mu^2} - i\pi \right)
    + \frac32\,\gamma_{\rm cusp}^{(3)}\,\ln\frac{m_G^2}{\mu^2} + \gamma^{G'} 
    + \{ \bm{\gamma}^{Q_L},\,.\, \} \,, 
    \notag\\
   \bm{\Gamma}_{q_R q_R}(\mu) 
   &= \left( e_q^2\,\gamma_{\rm cusp}^{(1)} - \frac16\,\gamma_{\rm cusp}^{(3)} \right) 
    \left( \ln\frac{m_G^2}{\mu^2} - i\pi \right)
    + \frac32\,\gamma_{\rm cusp}^{(3)}\,\ln\frac{m_G^2}{\mu^2} + \gamma^{G'} 
    + \{ \bm{\gamma}^{q_R},\,.\, \} \,; 
    \quad q=u,d , \notag\\
   \bm{\Gamma}_{GB}(\mu) 
   &= \frac32\,\gamma_{\rm cusp}^{(3)}\,\ln\frac{m_G^2}{\mu^2} + \gamma^{G'} + \gamma^G + \gamma^B \,.
\end{align}
The cusp terms without imaginary parts arise from soft gluon exchanges between the initial-state heavy resonance and one of the final-state particles. The single-particle anomalous dimensions of the gauge fields vanish at one-loop order \cite{Becher:2009cu,Becher:2009qa}, while the one-loop anomalous dimension of the heavy color-octet vector field is given by \cite{Becher:2009kw}
\begin{align}
   \gamma^{G'} = - \frac{3\alpha_3}{2\pi} + \dots \,.
\end{align}
Given these expressions, the resummation of large double and single logarithms can be accomplished in the same way as discussed in Section~\ref{sec:RGE}.

\renewcommand{\theequation}{6.\arabic{equation}}
\setcounter{equation}{0}

\section{Matching calculations for a UV completion of the SM}
\label{sec:matching}

In this section we illustrate our approach in the context of a concrete new-physics scenario, which has been proposed to address some of the $B$-meson anomalies observed in the transitions $b\to s\mu^+\mu^-$ \cite{Altmannshofer:2014cfa}. The setup we consider is an extension of the SM with a gauged $L_\mu-L_\tau$ lepton number, denoted by $U(1)'$. The new symmetry is broken spontaneously by the VEV $\langle S\rangle \equiv u/\sqrt{2}$ of a complex scalar field $S$, which is charged under the $U(1)'$ with $Q_S'=1$ and transforms as a singlet under the SM gauge group. The associated gauge boson $Z'$ acquires a mass $m_{Z'}=g_{Z'}\spac u$, where $g_{Z'}$ is the gauge coupling of the $U(1)'$ and we assume that $u\gg v$. The muon and tau lepton carry charges $+1$ and $-1$ under the $U(1)'$, whereas the remaining SM fermions are uncharged.

The model is supplemented by a single generation of vector-like quark (VLQ) partners $\mathbb{Q}$, $\mathbb{U}$ and $\mathbb{D}$, which transform like $Q_L$, $u_R$ and $d_R$ under the SM gauge group, but in addition are charged under the new $U(1)'$, i.e.\
\begin{align}
   \mathbb{Q} \sim (3,2)_{\frac16,\,1} \,, \qquad
   \mathbb{U} \sim (3,1)_{\frac23,\,-1} \,, \qquad
   \mathbb{D} \sim (3,1)_{-\frac13,\,-1} \,.
\end{align}
Being vector-like, these fermions are allowed to have masses without a Higgs-like mechanism,
\begin{align}\label{eq:Lm}
   \mathcal L_m = - m_Q\,\bar{\mathbb{Q}}\spac\mathbb{Q} - m_D\,\bar{\mathbb{D}}\spac\mathbb{D} 
    - m_U\,\bar{\mathbb{U}}\spac\mathbb{U} \,.
\end{align}
The scale of the VLQ mass terms $m_X$ (with $X=Q,U,D$) is not necessarily connected to the mass scale of the $Z'$ boson. Below, we consider the two cases where $m_X\sim u$ or $m_X\gg u$ in detail.

The chosen charges of the VLQs allow for the following Yukawa-type interactions with the SM quarks and the new scalar $S$:
\begin{align}\label{eq:LagVLQYuk}
   \mathcal L_Y = - \bar{\mathbb{Q}}\,\bm{Y}_Q^\dagger\,S\spac Q_L
    - \bar{\mathbb{U}}\,\bm{Y}_U^\dagger\,S^\dagger u_R - \bar{\mathbb{D}}\,\bm{Y}_D^\dagger\,S^\dagger d_R 
    + \mathrm{h.c.} \,. 
\end{align} 
Note that the new Yukawa couplings $\bm{Y}_X^\dagger$ are $1\times 3$ matrices in generation space, because we consider a single generation of VLQs. Upon the spontaneous breaking of the $U(1)'$ symmetry these interactions generate a mass mixing between the heavy VLQs and the massless SM fields, which results in induced couplings of the heavy $Z'$ boson to the SM quarks, with a non-trivial flavor structure. Flavor-conserving couplings of the $Z'$ boson to the SM fermions can also be generated through a kinetic mixing between the $Z'$ and the hypercharge gauge boson, as described by the operator
\begin{align}\label{Lmix}
   \mathcal L_\mathrm{mix} = - \frac{\xi}{2}\spac Z'_{\mu\nu}B^{\mu\nu} \,.
\end{align}
It is often assumed that the mixing parameter $\xi$ is one-loop suppressed. Finally, we note that a Higgs-portal interaction of the form $S^\dagger S\,\phi^\dagger\phi$ would not give any contributions to the matching conditions considered below and hence we do not need to discuss it in detail.

In this section we perform the matching of this renormalizable, anomaly-free extension of the SM \cite{Altmannshofer:2014cfa} onto the effective theory SCET$_{\rm BSM}$. Assuming that the $Z'$ boson will be the first new particle beyond the SM that is discovered experimentally, our setup provides a consistent framework for describing the two-body decays of the $Z'$ into SM particles in a model-independent and systematic way. We describe the matching procedure in detail for the Wilson coefficients $C_{\phi\phi}$ and $C_{\psi\psi}^{ij}$ appearing at leading order in the effective Lagrangian (\ref{eq:LagLambda2}). The fermionic operators are generated at tree level, while $O_{\phi\phi}$ occurs first at one-loop order (or via the kinetic mixing term). We first consider the case where the masses of the VLQs are of similar order as the mass of the $Z'$ boson. In Section~\ref{subsec:doubleh} we will then discuss the case of a double hierarchy $m_X\gg m_{Z'}\gg v$.

\subsection{Matching coefficients of the leading fermionic operators}

The tree-level matching contributions to the Wilson coefficients $C_{L_L L_L}^{ij}$ and $C_{e_R e_R}^{ij}$ describing the $Z'$ couplings to leptons are diagonal in the generation indices $i$ and $j$. By construction, the $Z'$ boson couples with different signs to the leptons of the second and third generation. In addition, the kinetic mixing in (\ref{Lmix}) gives a contribution proportional to the hypercharge of the various fermions. In matrix notation, we find
\begin{align}\label{leptoncoefs}
\begin{aligned}
   \bm{C}_{L_L L_L} 
   &= g_{Z'} \left( {\small \begin{array}{ccccr} 0 &&& 0 & 0 \\[-1mm] 0 &&& 1 & 0 \\[-1mm] 0 &&& 0 & -1 
    \end{array} } \right) + \frac{\xi\spac g'}{2}\,\bm{1} \,, \\
   \bm{C}_{e_R e_R} 
   &= g_{Z'} \left( {\small \begin{array}{ccccr} 0 &&& 0 & 0 \\[-1mm] 0 &&& 1 & 0 \\[-1mm] 0 &&& 0 & -1 
    \end{array} } \right) + \xi\spac g'\,\bm{1} \,.
\end{aligned}
\end{align}
Note that by virtue of $SU(2)_L$ invariance the operator $O_{LL}^{ij}$ also describes $Z'$-boson couplings to the neutrinos $\nu_\mu$ and $\nu_\tau$.

The corresponding couplings to the SM quarks have a more interesting flavor structure. To find them, we first need to diagonalize the quark mass terms in the Lagrangian. This is discussed in detail in Appendix~\ref{app:B}. Using the results derived there, it is straightforward to evaluate the matching conditions for the Wilson coefficients of the  operators coupling the $Z'$ boson to a pair of SM quarks. We find
\begin{align}\label{quarkcoefs}
\begin{aligned}
   \bm{C}_{Q_L Q_L} 
   &= \phantom{-} g_{Z'}\,\frac{u^2}{2 M_Q^2}\,\bm{Y}_Q \bm{Y}_Q^\dagger 
    - \frac{\xi\spac g'}{6}\,\bm{1} \,, \\
   \bm{C}_{u_R u_R} 
   &= - g_{Z'}\,\frac{u^2}{2 M_U^2}\,\bm{Y}_U \bm{Y}_U^\dagger 
    - \frac{2\spac\xi\spac g'}{3}\,\bm{1} \,, \\
   \bm{C}_{d_R d_R} 
   &= - g_{Z'}\,\frac{u^2}{2 M_D^2}\,\bm{Y}_D \bm{Y}_D^\dagger 
    + \frac{\xi\spac g'}{3}\,\bm{1} \,.
\end{aligned}
\end{align}

The matching coefficients in (\ref{leptoncoefs}) and (\ref{quarkcoefs}) are obtained at a factorization scale $\mu\sim M_X\sim m_{Z'}$. They can be evolved down to lower scales using the RG evolution equations derived in Section~\ref{sec:RGE}. 

\subsection{Matching coefficient of the leading bosonic operator}
\label{sec:matchPhiPhi}

\begin{figure}
\centering
\includegraphics[scale=0.75]{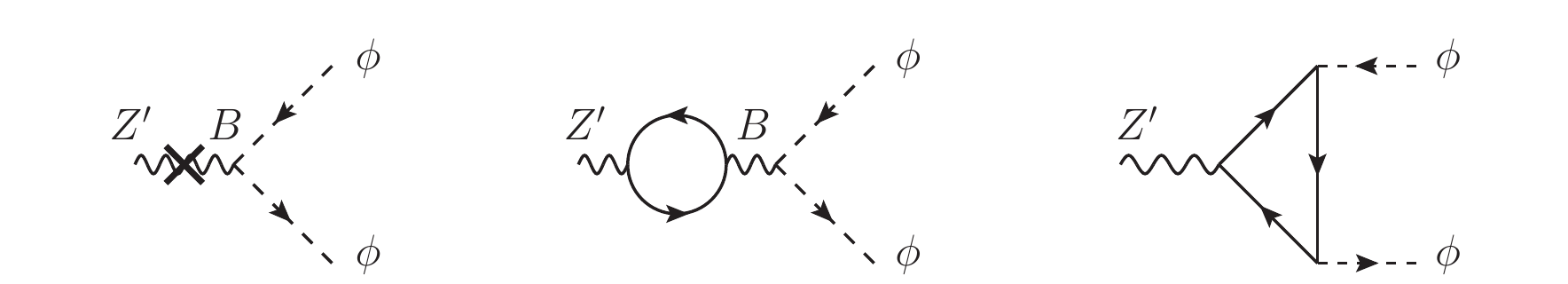}
\caption{\label{fig:Zphiphi} 
Representative Feynman diagrams contributing to the matching calculation for the operator $O_{\phi\phi}$. The cross in the first graph denotes an insertion of the kinetic-mixing interaction~(\ref{Lmix}).}
\end{figure}

Except for a contribution from the kinetic-mixing Lagrangian in (\ref{Lmix}), the bosonic operator $O_{\phi\phi}$ is generated first at the one-loop-level in the model we consider. To perform the matching calculation for the coefficient $C_{\phi\phi}$, we compute the amplitude $Z'\to\phi\phi^\ast$ in the unbroken phase of the SM. There exist three types of contributions, depicted in Figure~\ref{fig:Zphiphi}. We assume the kinetic-mixing parameter $\xi$ to be small, such that the first diagram is of the same order as the remaining one-loop graphs.

In order to avoid unnecessarily complicated expressions, we make the following simplifying assumptions about the model parameters: Without loss of generality we work in a basis where the up-quark Yukawa matrix $\bm{y}_u$ is diagonal. In this basis, we assume that the new Yukawa interactions coupling the VLQs to the SM quarks in (\ref{eq:LagVLQYuk}) only affect the third generation, and that the new Yukawa couplings are identical and real, i.e.\
\begin{align}
   \mathbf{Y}_Q^\dagger = \mathbf{Y}_U^\dagger = \mathbf{Y}_D^\dagger \equiv (0,0,Y_X) \,.
\end{align}
Also, we set the mass parameters $m_X$ of the VLQs equal to each other, so that the physical masses of the VLQs are identical and given by
\begin{align}
   M_X \equiv M_Q = M_U = M_D = \sqrt{m_X^2 + \frac{u^2}{2}\,Y_X^2} \,.
\end{align}
Finally, we neglect all SM Yukawa couplings with the exception of $(\bm{y}_u)_{33}=y_t$. Under these assumptions, the one-loop matching calculation can be straightforwardly performed, yielding
\begin{align}\label{eq:cphiphi}
   C_{\phi\phi} 
   = - \frac{\xi\spac g'}{2} + g_{Z'}\,\frac{3y_t^2}{8\pi^2}\,\frac{u^2\spac Y_X^2}{m_{Z'}^2}\,
    F\bigg(\frac{m_{Z'}^2}{M_X^2},\frac{u^2\spac Y_X^2}{2 M_X^2}\bigg) \,, 
\end{align}
where 
\begin{align}
   F(x,y) 
   &= (1-y)\,\bigg[ 1 - \frac{2(2-x y)}{x}\,\arcsin^2\frac{\sqrt{x}}{2}
    + (x+2y)\,\sqrt{\frac{4}{x}-1}\,\arcsin\frac{\sqrt{x}}{2} \notag\\
   &\hspace{2.0cm}\mbox{}- \frac{y(1+x)}{x}\,\mathrm{Li}_2(-x)
    - \frac{y(1+2x)}{x}\,\mathrm{Li}_2(x) + \frac{(x+2y)(1-x)}{x} \ln(1-x) \notag\\
    &\hspace{2.0cm}\mbox{}+ \left( \frac{x}{2} + y - \frac{y(1+x)}{x}\,\ln(1+x) 
     \right) \left( \ln x-i\pi \right) \bigg] 
     + 2\,\sqrt{\frac{4}{x}-1}\,\arcsin\frac{\sqrt{x}}{2} - 2 \,.
\end{align}
This expression exhibits branch cuts starting at $x=1$ (corresponding to $m_{Z'}=M_X$) and $x=4$ (corresponding to $m_{Z'}=2M_X$), and the prescription $x\equiv x+i0$ ensures that it is evaluated at the right side of these cuts. In (\ref{eq:cphiphi}), the first term accounts for the kinetic-mixing contribution. Note that the amplitude from the second diagram in Figure~\ref{fig:Zphiphi} vanishes under our assumptions due to the fact that the VLQs have degenerate masses and due to their specific charge assignments under the $U(1)'$ symmetry. In the third term, the coupling $g_{Z'}$ comes from the $Z'$ coupling to fermions, the top-quark Yukawa couplings arise from the two Higgs vertices, and the factor $u^2\spac Y_X^2$ arises from the mixing of the VLQs with the SM fermions.

Our result for the matching coefficient $C_{\phi\phi}$ is obtained at a high factorization scale $\mu\sim M_X$ and can be evolved down to lower scales using the RG evolution equations derived in Section~\ref{sec:RGE}. We emphasize the crucial fact that in SCET$_{\rm BSM}$ the matching coefficients depend on both, the mass $m_{Z'}$ of the heavy particle for whose decays the effective theory has been constructed, and the mass parameters of other heavy particles that are integrated out (the masses of the VLQs represented by $M_X$, and the VEV $u$ setting the mass of the scalar field $S$). 

\subsection{Matching and running in the double-hierarchy scenario}
\label{subsec:doubleh}

It is interesting to study the case in which the mass scale $M_X$ of the VLQs is much higher than $u\sim m_{Z'}$. It can be seen from (\ref{eq:cphiphi}) that the matching coefficients depend on two hierarchical scales in this case. Indeed, an expansion in the ratio $m_{Z'}^2/M_X^2\ll 1$ yields
\begin{align}\label{eq:94}
   C_{\phi\phi} 
   = - \frac{\xi\spac g'}{2} - g_{Z'}\,\frac{3y_t^2}{16\pi^2}\,\frac{u^2\spac Y_X^2}{M_X^2}
    \left( \ln\frac{M_X^2}{m_{Z'}^2} + i\pi + \frac12 \right) 
    + {\cal O}\bigg(\frac{u^4}{M_X^4}\bigg) \,. 
\end{align}
In order to obtain reliable predictions in perturbation theory, we should then perform two matching steps. At the scale $\mu\sim M_X$ we integrate out the VLQs and match the full theory onto an intermediate (local) EFT, comprised of the SM degrees of freedom as well as the $Z'$ boson and the scalar $S$. At the lower scale $\mu\sim m_{Z'}$ this EFT is then matched onto the SCET$_\mathrm{BSM}$. The RG evolution of the intermediate EFT between the scales $M_X$ and $m_{Z'}$ allows us to resum the large logarithms $\ln(M_X^2/m_{Z'}^2)$ to all orders in perturbation theory.

We begin by defining a basis of dimension-6 operators for the intermediate EFT, including only the operators relevant to our discussion. The corresponding effective Lagrangian is 
\begin{align}\label{eq:LEFTInter}
\begin{aligned}
   \mathcal L_\mathrm{eff} 
   &= \big( S^\dagger\spac i\!\overleftrightarrow{D}_{\!\mu}\spac S \big)\,\bigg[
    C^{ij}_Q\,\bar Q_L^i \gamma^\mu Q_L^j + C_u^{ij}\,\bar u_R^i \gamma^\mu u_R^j 
    + C_d^{ij}\,\bar d_R^i \gamma^\mu d_R^j 
    + C_\phi \big( \phi^\dagger\spac i\!\overleftrightarrow{D}_{\!\mu}\spac \phi \big) \bigg] \\
   &\quad\mbox{}- C_B\,S^\dagger S\,Z'_{\mu\nu}\spac B^{\mu\nu} + \dots \,,
\end{aligned} 
\end{align}
where the dots refer to operators that are irrelevant to our discussion. The Wilson coefficients $C^{ij}_Q$, $C^{ij}_u$ and $C^{ij}_d$ form the entries of $3\times 3$ hermitian matrices in generation space, while the coefficient $C_B$ is real. In addition to these operators there is still the kinetic-mixing Lagrangian shown in (\ref{Lmix}), but now with $\xi$ replaced by an effective value
\begin{align}\label{ximatch}
   \xi_{\rm eff} 
   = \xi - \frac{g_{Z'}\spac g'}{12\pi^2}\,\ln\frac{m_Q^2\spac m_D^2}{m_U^4} \,.
\end{align}
The matching contribution arising when one integrates out the VLQs is UV finite due to the chosen $U(1)'$ charges. Our result differs by a factor $(-6)$ from the corresponding expression given in Ref.~\cite{Altmannshofer:2014cfa}.

\begin{figure}
\centering
\includegraphics[scale=0.75]{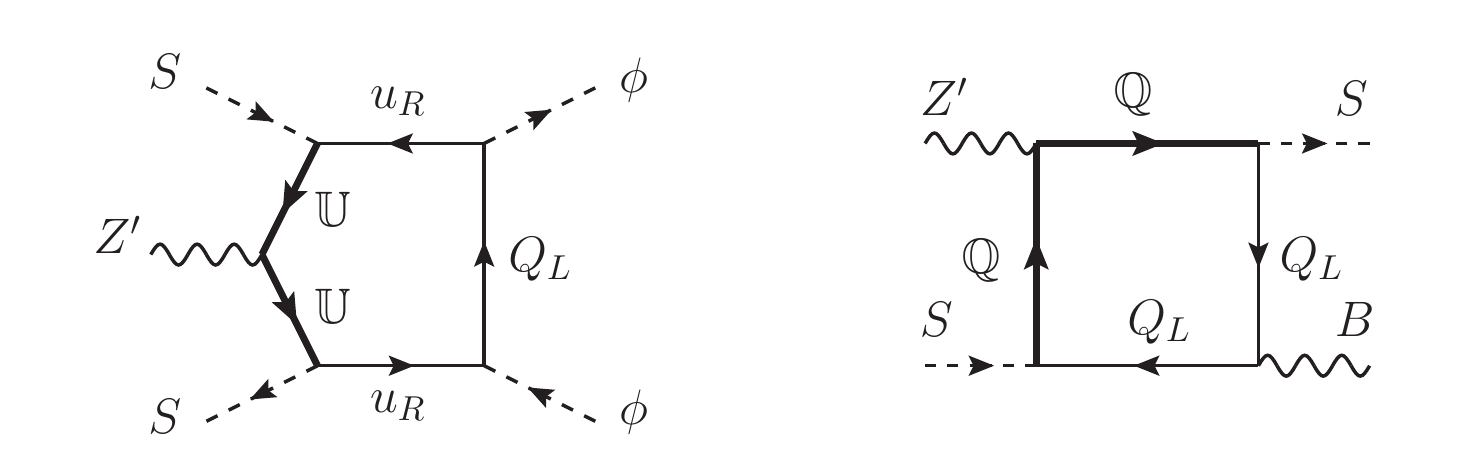}
\caption{\label{fig:CphiCB} 
Representative Feynman diagrams contributing to the one-loop matching conditions for the Wilson coefficients $C_\phi$ (left) and $C_B$ (right).}
\end{figure}

The first three operators in the first line of (\ref{eq:LEFTInter}) follow from integrating out the VLQs at tree level, which yields the coefficients (in matrix notation) \cite{Altmannshofer:2014cfa}
\begin{align}\label{eq:CfIntEFT}
   \bm{C}_Q = \frac{\bm{Y}_Q \bm{Y}_Q^\dagger}{2 m_Q^2} \,, \qquad 
   \bm{C}_u = - \frac{\bm{Y}_U \bm{Y}_U^\dagger}{2 m_U^2} \,, \qquad
   \bm{C}_d = - \frac{\bm{Y}_D \bm{Y}_D^\dagger}{2 m_D^2} \,.
\end{align}
The remaining operators arise first at one-loop order. Two representative Feynman graphs are depicted in Figure~\ref{fig:CphiCB}. For the coefficient $C_\phi$ we obtain 
\begin{align}\label{eq:CPhi1}
\begin{aligned}
   C_\phi 
   &= \frac{3}{16\pi^2}\,\bigg[
    \frac{\bm{Y}_Q^\dagger\spac\big( \bm{y}_u\spac\bm{y}_u^\dagger 
    - \bm{y}_d\,\bm{y}_d^\dagger \big) \bm{Y}_Q}{2m_Q^2}\,
    \bigg(\! \ln\frac{\mu^2}{m_Q^2} + \frac32 \bigg) \\
   &\hspace{1.7cm}\mbox{}+ \frac{\bm{Y}_U^\dagger\,\bm{y}_u^\dagger\spac\bm{y}_u \bm{Y}_U}{2m_U^2}\,
    \bigg(\! \ln\frac{\mu^2}{m_U^2} + \frac32 \bigg)
    - \frac{\bm{Y}_D^\dagger\,\bm{y}_d^\dagger\,\bm{y}_d\spac\bm{Y}_D}{2m_D^2}\, 
    \bigg(\! \ln\frac{\mu^2}{m_D^2} + \frac32 \bigg) \bigg] \,.
\end{aligned} 
\end{align}
The IR divergences of the relevant matching diagrams give rise to the logarithmic dependence on the factorization scale $\mu$ (after operator renormalization in the $\overline{\rm MS}$ scheme). Next we focus on the operator multiplying the Wilson coefficient $C_B$ in (\ref{eq:LEFTInter}), which is similar to the kinetic-mixing operator in (\ref{Lmix}). Indeed, when the scalar field $S$ acquires a VEV, this term gives rise to a power correction of order $u^2/m_X^2$ to the kinetic-mixing parameter $\xi$. Evaluating the relevant one-loop diagrams containing both VLQs and SM quarks in the loop, we find 
\begin{align}\label{CBres}
   C_B = \frac{g_{Z'} g'}{48\pi^2} \left[ 
    2\,\frac{\bm{Y}_U^\dagger \bm{Y}_U}{m_U^2}\,\bigg( \ln\frac{\mu^2}{m_U^2} + 2 \bigg) 
    - \frac{\bm{Y}_D^\dagger \bm{Y}_D}{m_D^2}\,\bigg( \ln\frac{\mu^2}{m_D^2} + 2 \bigg) 
    - \frac{\bm{Y}_Q^\dagger \bm{Y}_Q}{m_Q^2}\,\bigg( \ln\frac{\mu^2}{m_Q^2} + 2 \bigg) \right] .
\end{align}
Note that in the simplified scenario considered in Section~\ref{sec:matchPhiPhi} both $C_B$ and the matching contribution to $\xi_{\rm eff}$ in (\ref{ximatch}) vanish due to the fact that $m_Q=m_U=m_D$ and $\bm{Y}_Q=\bm{Y}_U=\bm{Y}_D$.

At the scale $\mu\sim m_{Z'}$ the intermediate EFT is matched onto SCET$_{\rm BSM}$. For the fermionic operators this matching is trivial at lowest order, since all we need to do is replace the scalar field $S$ in (\ref{eq:LEFTInter}) by its VEV and replace the various SM fields by their SCET counterparts. Including also the contributions from kinetic mixing, we find
\begin{align}
\begin{aligned}
   \bm{C}_{Q_L Q_L} 
   &= u^2\spac g_{Z'}\,\bm{C}_Q 
    - \big( \xi_{\rm eff} + u^2\spac C_B \big)\,\frac{g'}{6}\,\bm{1} \,, \\
   \bm{C}_{u_R u_R}
   &= u^2\spac g_{Z'}\,\bm{C}_u
    - \big( \xi_{\rm eff} + u^2\spac C_B \big)\,\frac{2g'}{3}\,\bm{1} \,, \\
   \bm{C}_{d_R d_R} 
   &= u^2\spac g_{Z'}\,\bm{C}_d 
    + \big( \xi_{\rm eff} + u^2\spac C_B \big)\,\frac{g'}{3}\,\bm{1} \,.
\end{aligned}
\end{align}
The results are consistent with those shown in (\ref{quarkcoefs}). The difference between $m_X^2$ and $M_X^2$ in the first terms amounts to power corrections of order $u^2/m_X^2$, which are neglected in the intermediate EFT. The higher-order contributions to the kinetic-mixing parameter $\xi$ arise from loop effects, which have been neglected in (\ref{quarkcoefs}). 

\begin{figure}
\centering
\includegraphics[scale=0.75]{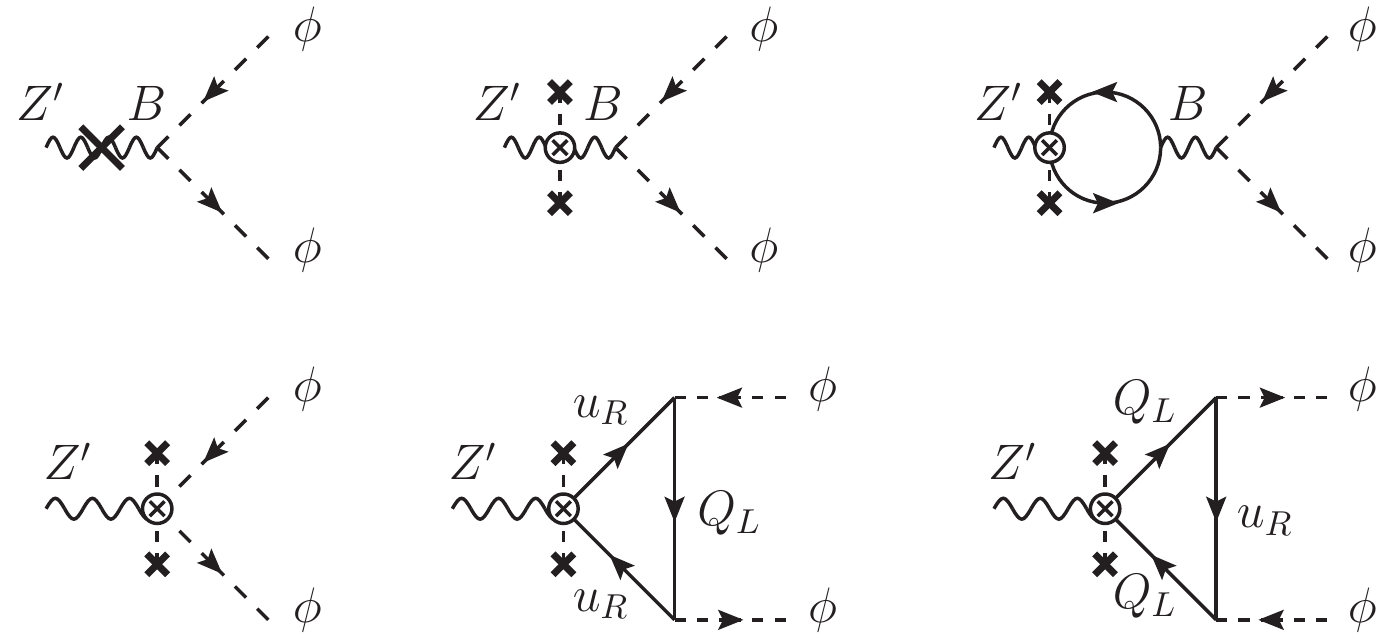} 
\caption{\label{fig:PhiPhiEFTIntermediate} 
Feynman diagrams contributing to the matching condition for the Wilson coefficient $C_{\phi\phi}$ in the double-hierarchy scenario. The crossed circles denote operator insertions of $O_Q^{ij}$, $O_u^{ij}$, $O_d^{ij}$, $O_\phi$ and $O_B$, corresponding to the dimension-6 operators shown in (\ref{eq:LEFTInter}). Dashed lines ending in crosses represent insertions of $\langle S \rangle$.}
\end{figure}

The matching condition for the Wilson coefficient of the bosonic operator $O_{\phi\phi}$, whose explicit form has been shown in (\ref{eq:94}), is more interesting. The diagrams contributing to this calculation are shown in Figure~\ref{fig:PhiPhiEFTIntermediate}. Contrary to the computation in Section~\ref{sec:matchPhiPhi}, the result in the two-step matching procedure can easily be obtained without simplifying the flavor structure of the model. We find
\begin{align}\label{eq:CPhiPhiDoubleHierarchy1}
\begin{aligned}
   C_{\phi\phi} 
   &= - \frac{g'}{2}\,\big( \xi_{\rm eff} + u^2\spac C_B \big)
   - \frac{g_{Z'} g^{\prime\,2} u^2}{48\pi^2}\,
   \mathrm{Tr}\big( 2\spac\bm{C}_u - \bm{C}_d + \bm{C}_Q \big)
    \left( \ln\frac{\mu^2}{m_{Z'}^2} + i\pi + \frac53 \right) \\
   &\quad\mbox{}+ g_{Z'}\spac u^2\,\bigg\{ C_\phi + \frac{3}{16\pi^2}\,\mathrm{Tr}\Big[ 
    \bm{y}_u^\dagger\spac\bm{y}_u\spac\bm{C}_u - \bm{y}_d^\dagger\,\bm{y}_d\,\bm{C}_d
    \!-\! \big( \bm{y}_u\spac\bm{y}_u^\dagger - \bm{y}_d\,\bm{y}_d^\dagger \big)\spac\bm{C}_Q \Big]\! 
    \left( \ln\frac{\mu^2}{m_{Z'}^2} + i\pi + 2 \right) \!\!\bigg\} \,,
\end{aligned}
\end{align}
where for simplicity we have included only the lowest-order contribution for each operator in the intermediate EFT. Note that the explicit scale dependence on the right-hand side of this equation cancels out when one inserts the one-loop expressions for the various Wilson coefficients from above. This leads to
\begin{align}
   C_{\phi\phi} 
   &= - \frac{g'}{2}\,\xi_{\rm eff} 
    + \frac{g_{Z'} g^{\prime\,2} u^2}{48\pi^2}\,\bigg[
    \frac{\bm{Y}_U^\dagger \bm{Y}_U}{m_U^2} 
    \left( \ln\frac{m_U^2}{m_{Z'}^2} + i\pi - \frac13 \right)
    - \frac{\bm{Y}_D^\dagger \bm{Y}_D}{2m_D^2} 
    \left( \ln\frac{m_D^2}{m_{Z'}^2} + i\pi - \frac13 \right) \notag\\
   &\hspace{4.2cm}\mbox{}- \frac{\bm{Y}_Q^\dagger \bm{Y}_Q}{2m_Q^2} 
    \left( \ln\frac{m_Q^2}{m_{Z'}^2} + i\pi - \frac13 \right) \!\bigg] \notag\\
   &\quad\mbox{}- g_{Z'}\spac \frac{3 u^2}{16\pi^2}\,\bigg[
    \frac{\bm{Y}_U^\dagger\,\bm{y}_u^\dagger\spac\bm{y}_u\spac\bm{Y}_U}{2m_U^2} 
    \left( \ln\frac{m_U^2}{m_{Z'}^2} + i\pi + \frac12 \right)
    - \frac{\bm{Y}_D^\dagger\,\bm{y}_d^\dagger\,\bm{y}_d\spac\bm{Y}_D}{2m_D^2} 
    \left( \ln\frac{m_D^2}{m_{Z'}^2} + i\pi + \frac12 \right) \notag\\
   &\hspace{4.2cm}\mbox{}+ \frac{\bm{Y}_Q^\dagger\spac\big( \bm{y}_u\spac\bm{y}_u^\dagger 
    - \bm{y}_d\,\bm{y}_d^\dagger \big)\spac\bm{Y}_Q}{2m_Q^2}
    \left( \ln\frac{m_Q^2}{m_{Z'}^2} + i\pi + \frac12 \right) \!\bigg] \,.
\end{align}
In the approximation adopted in Section~\ref{sec:matchPhiPhi}, where we have set $m_Q=m_U=m_D\equiv m_X$, $\bm{Y}_Q=\bm{Y}_U=\bm{Y}_D\equiv (0,0,Y_X)^T$ and neglected all SM Yukawa couplings other than $(\bm{y}_u)_{33}=y_t$, the above result reduces to the one shown in (\ref{eq:94}).

\begin{figure}
\centering
\includegraphics[scale=0.75]{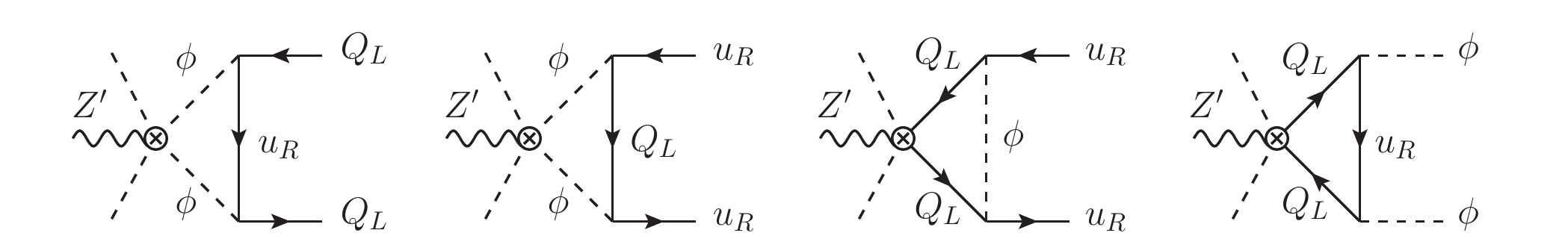}
\vspace{2mm}
\caption{\label{fig:RGEs} 
Representative Feynman diagrams contributing to the mixing of the operators $O_Q^{ij}$, $O_u^{ij}$, $O_d^{ij}$ and $O_\phi$. These graphs must be supplemented by external-leg corrections (not shown).}
\end{figure}

The two-scale matching procedure described here allows us to improve the above expression by resumming the large logarithms $\ln(m_X^2/m_{Z'}^2)$, where $X=Q,u,d$, to all orders in perturbation theory. To this end, we evaluate the Wilson coefficients in the intermediate EFT, given at lowest order in \eqref{eq:CfIntEFT}--\eqref{CBres}, at a scale $\mu_X\sim m_X$, where they are free of large logarithms. Here $m_X\sim m_Q\sim m_U\sim m_D$ is the characteristic mass scale of the heavy VLQs. We then evolve these coefficients to a scale $\mu_{Z'}\sim m_{Z'}$ by solving their RG evolution equations. Finally, we insert the evolved coefficients $C_{B,\phi}(\mu_{Z'})$ and $\bm{C}_{Q,u,d}(\mu_{Z'})$ into relation (\ref{eq:CPhiPhiDoubleHierarchy1}), which for $\mu\sim m_{Z'}$ is also free of large logarithms. We have calculated the one-loop anomalous dimensions governing the scale evolution of the Wilson coefficients in the effective Lagrangian of the intermediate EFT. For the coefficients of the operators shown in the first line of (\ref{eq:LEFTInter}) diagrams such as those in Figure~\ref{fig:RGEs} need to be computed. We find that only Yukawa interactions contribute at one-loop order, and that in matrix notation the resulting evolution equations take the form 
\begin{align}\label{eq:RGevol}
\begin{aligned}
   \frac{d}{d\ln\mu}\,\bm{C}_Q(\mu)
   &= \frac{1}{32\pi^2}\,\{ \bm{y}_u\spac\bm{y}_u^\dagger+\bm{y}_d\,\bm{y}_d^\dagger, \bm{C}_Q \}
    - \frac{1}{16\pi^2} \left( \bm{y}_u\spac\bm{C}_u\spac\bm{y}_u^\dagger
    + \bm{y}_d\,\bm{C}_d\,\bm{y}_d^\dagger \right) \\
   &\quad\mbox{}+ \frac{C_\phi}{16\pi^2} \left( \bm{y}_u\spac\bm{y}_u^\dagger
    - \bm{y}_d\,\bm{y}_d^\dagger \right) , \\
   \frac{d}{d\ln\mu}\,\bm{C}_u(\mu)
   &= \frac{1}{16\pi^2}\,\{ \bm{y}_u^\dagger\spac\bm{y}_u, \bm{C}_u \}
    - \frac{1}{8\pi^2}\,\bm{y}_u^\dagger\spac\bm{C}_Q\,\bm{y}_u 
    - \frac{C_\phi}{8\pi^2}\,\bm{y}_u^\dagger\spac\bm{y}_u \,, \\
   \frac{d}{d\ln\mu}\,\bm{C}_d(\mu)
   &= \frac{1}{16\pi^2}\,\{ \bm{y}_d^\dagger\,\bm{y}_d, \bm{C}_d \}
    - \frac{1}{8\pi^2}\,\bm{y}_d^\dagger\,\bm{C}_Q\,\bm{y}_d 
    + \frac{C_\phi}{8\pi^2}\,\bm{y}_d^\dagger\,\bm{y}_d \,, \\
   \frac{d}{d\ln\mu}\,C_\phi(\mu)
   &= \frac{3}{8\pi^2}\,\text{Tr}\Big[ 
    \bm{C}_Q \big( \bm{y}_u\spac\bm{y}_u^\dagger - \bm{y}_d\,\bm{y}_d^\dagger \big) \Big]
    + \frac{3}{8\pi^2}\,\text{Tr}\big( 
    \bm{C}_d\,\bm{y}_d^\dagger\,\bm{y}_d - \bm{C}_u\,\bm{y}_u^\dagger\spac\bm{y}_u \big) \\
   &\quad\mbox{}+ \frac{3}{8\pi^2}\,C_\phi\,\text{Tr}\big( \bm{y}_u\spac\bm{y}_u^\dagger
    + \bm{y}_d\,\bm{y}_d^\dagger \big) \,,
\end{aligned}
\end{align}
where all quantities on the right-hand side are evaluated at the scale $\mu$. Note that the operator $O_B$ does not mix into the remaining operators under renormalization, and hence there are no contributions proportional to the Wilson coefficient $C_B$ in these equations.\footnote{This statement remains true when a portal coupling of the form $S^\dagger S\,\phi^\dagger\phi$ is included in the effective Lagrangian of the intermetidate EFT.} 
The anomalous dimensions entering above can also be inferred from the results obtained in Ref.~\cite{MartinCamalich:2020dfe}. Compared with the findings of these authors, we obtain different signs for all terms proportional to $C_\phi$ on the right-hand side of (\ref{eq:RGevol}). 
To solve the system of equations (\ref{eq:RGevol}), we work in the basis where the up-quark Yukawa matrix is diagonal and neglect all Yukawa couplings other than $y_t$, which is an excellent approximation numerically. Expression (\ref{eq:CPhiPhiDoubleHierarchy1}) then reduces to
\begin{align}\label{eq6.21}
   C_{\phi\phi}(m_{Z'}) 
   &= - \frac{g'}{2}\,\Big[ \xi_{\rm eff} + u^2\spac C_B(m_{Z'}) \Big] \notag\\[1mm]
   &\quad\mbox{}- \frac{g_{Z'} g^{\prime\,2} u^2}{48\pi^2}\,
    \mathrm{Tr}\Big[ \bm{C}_Q(m_{Z'}) + 2\spac\bm{C}_u(m_{Z'}) - \bm{C}_d(m_{Z'}) \Big]
    \left( \frac53 + i\pi \right) \\[-1mm]
   &\quad\mbox{}+ g_{Z'}\spac u^2\,\bigg\{ C_\phi(m_{Z'}) 
    - \frac{3\spac y_t^2(m_{Z'})}{16\pi^2}\,\Big[ C_Q^{33}(m_{Z'}) - C_u^{33}(m_{Z'}) \Big] 
    \left( 2 + i\pi \right) \!\!\bigg\} \,, \notag
\end{align}
where $C_{u,Q}^{nn}\equiv(\bm{C}_{u,Q})_{nn}$, and we have chosen $\mu_{Z'}=m_{Z'}$ for simplicity. The relevant evolution equations simplify to (there is no need to consider off-diagonal indices in generation space)
\begin{align}
\begin{aligned}
   \frac{d}{d\ln\mu}\,C_X^{11}(\mu) 
   &= \frac{d}{d\ln\mu}\,C_X^{22}(\mu) = \frac{d}{d\ln\mu}\,C_d^{33}(\mu)= 0 \,; \quad X = Q,u,d \,, \\
   \frac{d}{d\ln\mu}\,C_Q^{33}(\mu) 
   &= \frac{y_t^2(\mu)}{16\pi^2}\,\Big[ C_Q^{33}(\mu) - C_u^{33}(\mu) + C_\phi(\mu) \Big] \,, \\
   \frac{d}{d\ln\mu}\,C_u^{33}(\mu)
   &= - \frac{y_t^2(\mu)}{8\pi^2}\,\Big[ C_Q^{33}(\mu) - C_u^{33}(\mu) + C_\phi(\mu) \Big] \,, \\
   \frac{d}{d\ln\mu}\,C_\phi(\mu)
   &= \frac{3\spac y_t^2(\mu)}{8\pi^2}\,\Big[ C_Q^{33}(\mu) - C_u^{33}(\mu) + C_\phi(\mu) \Big] \,.
\end{aligned}
\end{align}
The solution to this system of equations can be written in the form (with $X=Q,u,d$)
\begin{align}
   C_X^{11}(m_{Z'}) = C_X^{11}(m_X) \,, \qquad
   C_X^{22}(m_{Z'}) = C_X^{22}(m_X) \,, \qquad
   C_d^{33}(m_{Z'}) = C_d^{33}(m_X) \,, 
\end{align}
and 
\begin{align}
\begin{aligned}
   C_Q^{33}(m_{Z'}) &= C_Q^{33}(m_X) 
    + \frac19\,\Big[ C_Q^{33}(m_X) - C_u^{33}(m_X) + C_\phi(m_X) \Big]\,U(m_X,m_{Z'}) \,, \\
   C_u^{33}(m_{Z'}) &= C_u^{33}(m_X) 
    - \frac29\,\Big[ C_Q^{33}(m_X) - C_u^{33}(m_X) + C_\phi(m_X) \Big]\,U(m_X,m_{Z'}) \,, \\
   C_\phi(m_{Z'}) &= C_\phi(m_X) 
    + \frac23\,\Big[ C_Q^{33}(m_X) - C_u^{33}(m_X) + C_\phi(m_X) \Big]\,U(m_X,m_{Z'}) \,,
\end{aligned}
\end{align}
where
\begin{align}
   U(m_X,m_{Z'}) = \exp\left[\, \int_{m_X}^{m_{Z'}}\!\frac{d\mu}{\mu}\,
    \frac{9\spac y_t^2(\mu)}{16\pi^2} \right] - 1 \,.
\end{align}
In this solution the large logarithms $\ln(m_X^2/m_{Z'}^2)$ are resummed in the leading logarithmic approximation and to all orders of perturbation theory. In the leading logarithmic approximation, where the anomalous dimensions are computed at one-loop order, one should use the tree-level matching conditions in this solution. For the coefficients needed in (\ref{eq6.21}), this gives
\begin{align}\label{bluemagic}
\begin{aligned}
   \mathrm{Tr}\,\bm{C}_Q(m_{Z'}) &= \frac{\bm{Y}_Q^\dagger \bm{Y}_Q}{2m_Q^2}
    + \frac{1}{18}\,\bigg( \frac{|Y_{Q,3}|^2}{m_Q^2} + \frac{|Y_{U,3}|^2}{m_U^2} \bigg)\,
    U(m_X,m_{Z'})  \,, \\
   \mathrm{Tr}\,\bm{C}_u(m_{Z'}) &= - \frac{\bm{Y}_U^\dagger \bm{Y}_U}{2m_U^2} 
    - \frac19\,\bigg( \frac{|Y_{Q,3}|^2}{m_Q^2} + \frac{|Y_{U,3}|^2}{m_U^2} \bigg)\,
    U(m_X,m_{Z'})  \,, \\
   \mathrm{Tr}\,\bm{C}_d(m_{Z'}) &= - \frac{\bm{Y}_D^\dagger \bm{Y}_D}{2m_D^2} \,, \\
   C_Q^{33}(m_{Z'}) - C_u^{33}(m_{Z'}) 
   &= \bigg( \frac{|Y_{Q,3}|^2}{m_Q^2} + \frac{|Y_{U,3}|^2}{m_U^2} \bigg)
     \left[ \frac12 + \frac16\,U(m_X,m_{Z'}) \right] \,, \\
   C_\phi(m_{Z'}) &= \frac13\,\bigg( \frac{|Y_{Q,3}|^2}{m_Q^2} + \frac{|Y_{U,3}|^2}{m_U^2} \bigg)\,
    U(m_X,m_{Z'}) \,,
\end{aligned}
\end{align}

\begin{figure}
\centering
\includegraphics[scale=0.75]{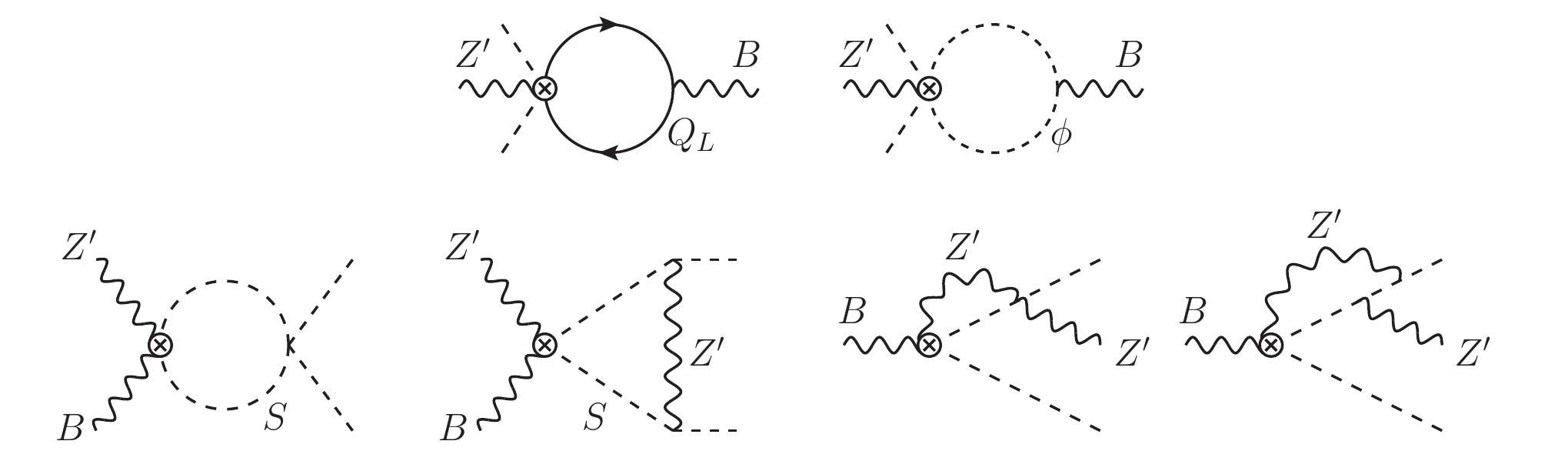}
\vspace{2mm}
\caption{\label{fig:RGECB} 
Representative Feynman diagrams contributing to the mixing of the operators $O_{Q,u,d}^{ij}$ and $O_\phi$ into $O_B$ (top row) and to the multiplicative renormalization of $O_B$ (bottom row). These graphs must be supplemented by external-leg corrections (not shown).}
\end{figure}

The evolution equation for the Wilson coefficient $C_B$ of the operator shown in the second line of (\ref{eq:LEFTInter}) receives contributions from all of the remaining operators. The relevant Feynman graphs are illustrated in Figure~\ref{fig:RGECB}. We obtain 
\begin{align}
   \frac{d}{d\ln\mu}\,C_B(\mu)
   = \gamma_B\,C_B - \frac{g_{Z'} g'}{12\pi^2}\,\text{Tr}\big( \bm{C}_Q + 2\bm{C}_u - \bm{C}_d \big)
    - \frac{g_{Z'} g'}{24\pi^2}\,C_\phi \,,
\end{align}
with
\begin{align}
   \gamma_B = \frac{7 g_{Z'}^2}{48\pi^2} + \frac{41 g^{\prime\,2}}{96\pi^2}
    + \frac{\lambda_S}{8\pi^2} \,,
\end{align}
where again all quantities on the right-hand side are evaluated at the scale $\mu$. The quartic coupling $\lambda_S$ is defined as ${\cal L}_{\rm quartic}=-\frac{\lambda_S}{4}\,(S^\dagger S)^2$. The general solution of this equation reads
\begin{align}\label{CBsol}
\begin{aligned}
   C_B(m_{Z'}) &= C_B(m_X)\,U_B(m_X,m_{Z'}) \\
   &\quad\mbox{}- \int_{m_X}^{m_{Z'}}\!\frac{d\mu}{\mu}\,\frac{g_{Z'}(\mu)\spac g'(\mu)}{12\pi^2}
    \left[ \text{Tr}\Big[ \bm{C}_Q(\mu) + 2\bm{C}_u(\mu) - \bm{C}_d(\mu) \Big]
    + \frac{C_\phi(\mu)}{2} \right] U_B(\mu,m_{Z'}) \,,
\end{aligned}
\end{align}
\vspace{-1mm}
where 
\begin{align}
   U_B(m_X,m_{Z'}) = \exp\left[\, \int_{m_X}^{m_{Z'}}\!\frac{d\mu}{\mu}\,\gamma_B(\mu) \right] .
\end{align}
In the leading logarithmic approximation one should use the tree-level matching condition $C_B(m_X)=0$ along with the solutions (\ref{bluemagic}) when evaluating expression (\ref{CBsol}). At one-loop order, the running coupling $g_{Z'}(\mu)$ satisfies the evolution equation
\begin{align}
   \frac{d\spac g_{Z'}}{d\ln\mu} = \frac{13\spac g_{Z'}^3}{48\pi^2} + \dots \,.
\end{align}

\section{Conclusions}
\label{sec:concl}

We have constructed an effective field theory describing the decays into SM particles of a new heavy vector resonance with mass far above the electroweak scale. Our approach implements a consistent expansion of the corresponding decay rates in powers of the ratio of the electroweak scale and the resonance mass. It is completely model independent and allows for arbitrarily complicated UV completions, which can have multiple heavy particles beyond the vector resonance and additional sectors at scales higher than the resonance mass itself. The light SM particles are described by collinear and soft fields in the language of SCET, while the massive resonance is treated using a heavy-vector effective theory akin to HQET. The latter is necessary to provide a consistent power counting as well as to alleviate the issue of renormalizability in theories with massive vector bosons. For the important example of a heavy $Z'$ boson, which is a singlet under the SM gauge group, we have constructed the operator basis for two-body (or two-jet) decays into SM particles at leading and next-to-leading order in the power expansion, and we have expressed the corresponding decay rates in terms of the Wilson coefficients of these operators. We have also derived the RG evolution equations for the Wilson coefficients and solved them for a few representative cases. In this way, large logarithmic corrections to the decay rates can be resummed to all orders in perturbation theory, generally resulting in significant ${\cal O}(1)$ corrections. With the example of a massive color-octet vector resonance (e.g.\ a Kaluza-Klein gluon), we have discussed the extension of our framework to the case of non-singlet resonances. The formulation of the effective theory presented in this work is done in the symmetric phase of the SM. As a consequence, for processes in which the characteristic mass scales of the final states lie far below the electroweak scale, one would need to perform an additional matching step onto SCET$_{\rm BSM}$ operators defined in the broken phase, which are invariant under $SU(3)_c\times U(1)_{\rm em}$. This matching is straightforward, and hence we have not discussed it here in detail. 

To illustrate our approach with a concrete example, we have performed the matching procedure for the most relevant operators in a specific UV model, consisting of an extension of the SM by a massive $Z'$ boson of a new $U(1)'$ symmetry for a gauged $L_\mu-L_\tau$ lepton number, which is broken spontaneously by the VEV of a new scalar field. Via a set of heavy VLQs the $Z'$ boson also couples to the SM quarks. This model features two new-physics scales: the mass scale of the VLQs and the VEV of the scalar field, which sets the mass of the $Z'$ boson. If these two scales are very hierarchical, large logarithms of their ratio can be resummed by using a two-step matching procedure, which we have discussed in detail.

Our framework represents an economic way of obtaining compact analytic expressions for the decay rates of new massive vector resonances, including the potentially sizable effects from the resummation of double and single Sudakov logarithms. In contrast to performing this resummation using parton showers, as is often done in phenomenological studies, our approach does not require Monte Carlo tools but relies on solving simple differential equations. Importantly, this allows for the inclusion of effects beyond the capabilities of parton showers, including the resummation of electroweak logarithms and of logarithms arising from the Yukawa interactions. Finally, the resummation can straightforwardly be extended to higher orders in perturbation theory.

With the generalization to heavy particles with non-zero spin and resonances carrying non-trivial SM charges, the present work constitutes two important generalizations of the SCET$_{\rm BSM}$ approach, which was originally introduced to study the decays of a hypothetical spin-0 SM singlet resonance \cite{Alte:2018nbn,Alte:2019iug}. All that is missing now is the long-awaited discovery of some new heavy particle not contained in the SM of particle physics.

\subsubsection*{Acknowledgments}

M.N.~thanks Gino Isidori, the particle theory group at Zurich University and the Pauli Center for hospitality during a sabbatical stay. The research of M.H.\ and M.N.\ was supported by the Cluster of Excellence {\em Precision Physics, Fundamental Interactions and Structure of Matter\/} (PRISMA${}^+$ -- EXC~2118/1) within the German Excellence Strategy (project ID 39083149). M.K.~gratefully acknowledges support by the Swiss National Science Foundation (SNF) under contract 200021-175940 and the European Union's Horizon 2020 Research and Innovation Programme under grant agreement 833280 (FLAY).

\begin{appendix}

\renewcommand{\theequation}{A.\arabic{equation}}
\setcounter{equation}{0}

\section{Operator basis at subleading order}
\label{app:A}

Here we present a complete basis of ${\cal O}(\lambda^3)$ two-jet operators relevant to the decays of a heavy $Z'$ boson. Without loss of generality we set $x=0$ for the spacetime point at which these operators are evaluated. This means that we do not need to include operators containing coordinate vectors (such as insertions of $x_\perp\cdot\partial_\perp$) arising from the multipole expansion of soft fields, which can in principle arise in higher orders \cite{Beneke:2002ph}. The operators listed below are genuine power-suppressed SCET operators. In addition there are time-ordered products of the leading-order operators in (\ref{eq:LagLambda2}) with power-suppressed terms in the SCET or HVET Lagrangians.

\subsubsection*{Fermionic operators}

A basis of operators in which a $Z'$ boson couples to a pair of fermions with equal chiralities can be chosen as
\begin{align}
\begin{aligned}
   O_1^{ij} &= \Pi\cdot Z_v'\,\big( \bar\Psi_n^i\,\Asl_s^\perp\Psi_\nb^j
    - \bar\Psi_\nb^i\,\Asl_s^\perp\Psi_n^j \big) \,, \\
   O_2^{ij}(u) &= \Pi\cdot Z_v'\,\big( \bar\Psi_n^i\,\Asl_n^{(u)\perp}\Psi_\nb^j 
    - \bar\Psi_\nb^i\,\Asl_\nb^{(u)\perp}\Psi_n^j \big) \,, \\
   O_3^{ij}(u) &= g_{\mu\nu}^\perp\,Z_v^{\prime\mu} \big(
    \bar\Psi_n^{(u)i}\,\vsl\,\A_\nb^{\perp\nu}\,\Psi_n^j
    + \bar\Psi_\nb^{(u)i}\,\vsl\,\A_n^{\perp\nu}\,\Psi_\nb^j \big) \,, \\
   O_4^{ij}(u) &= \epsilon_{\mu\nu}^\perp\,Z_v^{\prime\mu} \big(
    \bar\Psi_n^{(u)i}\,\vsl\,\A_\nb^{\perp\nu}\,\Psi_n^j
    - \bar\Psi_\nb^{(u)i}\,\vsl\,\A_n^{\perp\nu}\,\Psi_\nb^j \big) \,, \\
   \colored{ O_5^{ij} } &= \colored{ 
    (i\partial_\mu\spac\Pi\cdot Z_v')\,\big( \bar\Psi_n^i\gamma_\perp^\mu\Psi_\nb^j
    - \bar\Psi_\nb^i\gamma_\perp^\mu\Psi_n^j \big) } \,, \\
   \colored{ O_6^{ij} } &= \colored{ 
    \Pi\cdot Z_v'\,\big( \bar\Psi_n^i\,i\!\overleftrightarrow{\delsl}\!\!_\perp\Psi_\nb^j 
    - \bar\Psi_\nb^i\,i\!\overleftrightarrow{\delsl}\!\!_\perp\Psi_n^j \big) } \,.
\end{aligned}
\end{align}
Here and below, operators shown in gray can be omitted from the basis. Since the $Z'$ boson is a gauge singlet and hence does not interact, the matrix elements of operators including a transverse derivative on the field $Z_v^{\prime\mu}$, such as $O_5^{ij}$, are proportional to the transverse momentum of the $Z'$ boson, and one can always choose a reference frame in which this transverse momentum vanishes. The operator $O_6^{ij}$ can be eliminated using the equations of motion for the fermion fields. When an operator contains more than one collinear field in the same sector, these fields share the total collinear momentum in that sector. A variable $u\in[0,1]$ then indicates the fraction of the large component of the collinear momentum carried by one of the two fields, as indicated by the superscript ``$(u)$'' \cite{Alte:2018nbn}. The above operators are multiplied by corresponding Wilson coefficients, and in the effective Lagrangian one must take these products plus their hermitian conjugates. The Wilson coefficients of the operator $O_1^{ij}$ form the entries of a hermitian $3\times 3$ matrix in generation space, while the Wilson coefficients of $O_2^{ij}$ can be arbitrary complex numbers. The Wilson coefficients of the operators $O_{3,4}^{ij}$ satisfy the relations $C_{3,4}^{*ji}(u)=-C_{3,4}^{ij}(1-u)$.

A basis of operators in which a $Z'$ boson couples to a pair of fermions with opposite chiralities can be chosen as
\begin{align}
\begin{aligned}
   Q_1^{ij} &= \Pi\cdot Z_v'\,\big( \bar\Psi_{L,n}^i\Phi_0\Psi_{R,\nb}^j 
    - \bar\Psi_{L,\nb}^i\Phi_0\Psi_{R,n}^j \big) = m_{Z'}\,O_{\psi_L\psi_R}^{ij} \,, \\
   Q_2^{ij}(u) &= \Pi\cdot Z_v'\,\big( \bar\Psi_{L,n}^i\Phi_n^{(u)}\Psi_{R,\nb}^j 
    - \bar\Psi_{L,\nb}^i\Phi_\nb^{(u)}\Psi_{R,n}^j \big) \,, \\
   Q_3^{ij}(u) &= \Pi\cdot Z_v'\,\big( \bar\Psi_{L,n}^i\Phi_\nb^{(u)}\Psi_{R,\nb}^j 
    - \bar\Psi_{L,\nb}^i\Phi_n^{(u)}\Psi_{R,n}^j \big) \,, \\
   Q_4^{ij}(u) &= g_{\mu\nu}^\perp\,Z_v^{\prime\mu} \big(
    \bar\Psi_{L,n}^{(u)i}\,\Phi_\nb\,\vsl\,\gamma_\perp^\nu\Psi_{R,n}^j 
    + \bar\Psi_{L,\nb}^{(u)i}\,\Phi_n\,\vsl\,\gamma_\perp^\nu\Psi_{R,\nb}^j \big) \,.
\end{aligned}
\end{align}
The Wilson coefficients of these operators can be arbitrary complex numbers. There is no need to include an operator analogous to $Q_4^{ij}$ with $g_{\mu\nu}^\perp$ replaced by $\epsilon_{\mu\nu}^\perp$, because relation (\ref{Hillrela}) can be used to relate such an operator to $Q_4^{ij}$.

\subsubsection*{Operators containing two Higgs fields}

A basis of operators in which a $Z'$ boson couples to a pair of Higgs doublets can be chosen as
\begin{align}
\begin{aligned}
   P_1 &= g_{\mu\nu}^\perp\,Z_v^{\prime\mu}\,\big( 
    \Phi_n^\dagger\,\A_\nb^{\perp\nu} \Phi_0
    + \Phi_\nb^\dagger\,\A_n^{\perp\nu} \Phi_0 \big) = O_{A\phi}^\parallel \,, \\ 
   P_2 &= \epsilon_{\mu\nu}^\perp\,Z_v^{\prime\mu}\,\big( 
    \Phi_n^\dagger\,\A_\nb^{\perp\nu} \Phi_0
    - \Phi_\nb^\dagger\,\A_n^{\perp\nu} \Phi_0 \big) = O_{A\phi}^\perp \,, \\
   P_3 &= g_{\mu\nu}^\perp\,Z_v^{\prime\mu}\,\big( 
    \Phi_n^\dagger\,\A_\nb^{\perp\nu} \Phi_s
    + \Phi_\nb^\dagger\,\A_n^{\perp\nu} \Phi_s \big) \,, \\ 
   P_4 &= \epsilon_{\mu\nu}^\perp\,Z_v^{\prime\mu}\,\big( 
    \Phi_n^\dagger\,\A_\nb^{\perp\nu} \Phi_s
    - \Phi_\nb^\dagger\,\A_n^{\perp\nu} \Phi_s \big) \,, \\
   P_5 &= g_{\mu\nu}^\perp\,Z_v^{\prime\mu}\,\big( 
    \Phi_n^\dagger\,\A_s^\nu\,\Phi_\nb
    + \Phi_\nb^\dagger\,\A_s^\nu\,\Phi_n \big) \,, \\
   P_6 &= \epsilon_{\mu\nu}^\perp\,Z_v^{\prime\mu}\,\big( 
    \Phi_n^\dagger\,\A_s^\nu\,\Phi_\nb
    - \Phi_\nb^\dagger\,\A_s^\nu\,\Phi_n \big) \,, \\
   P_7 &= g_{\mu\nu}^\perp\,Z_v^{\prime\mu}\,\big( 
    \Phi_n^\dagger\,i\!\overleftrightarrow{\partial}\!\!_\perp^{\,\nu} \Phi_\nb
    + \Phi_\nb^\dagger\,i\!\overleftrightarrow{\partial}\!\!_\perp^{\,\nu} \Phi_n \big) \,, \\
   P_8 &= \epsilon_{\mu\nu}^\perp\,Z_v^{\prime\mu}\,\big( 
    \Phi_n^\dagger\,i\!\overleftrightarrow{\partial}\!\!_\perp^{\,\nu} \Phi_\nb
    - \Phi_\nb^\dagger\,i\!\overleftrightarrow{\partial}\!\!_\perp^{\,\nu} \Phi_n \big) \,, \\
   P_9(u) &= g_{\mu\nu}^\perp\,Z_v^{\prime\mu}\,\big( 
    \Phi_n^\dagger\,\A_n^{(u)\perp\nu} \Phi_\nb
    + \Phi_\nb^\dagger\,\A_\nb^{(u)\perp\nu} \Phi_n \big) \,, \\
   P_{10}(u) &= \epsilon_{\mu\nu}^\perp\,Z_v^{\prime\mu}\,\big( 
    \Phi_n^\dagger\,\A_n^{(u)\perp\nu} \Phi_\nb
    - \Phi_\nb^\dagger\,\A_\nb^{(u)\perp\nu} \Phi_n \big) \,, \\
   P_{11}(u) &= g_{\mu\nu}^\perp\,Z_v^{\prime\mu}\,\big( 
    \Phi_n^{(u)\dagger}\spac\A_\nb^{\perp\nu} \Phi_n
    + \Phi_\nb^{(u)\dagger}\spac\A_n^{\perp\nu} \Phi_\nb \big) \,, \\ 
   P_{12}(u) &= \epsilon_{\mu\nu}^\perp\,Z_v^{\prime\mu}\,\big( 
    \Phi_n^{(u)\dagger}\spac\A_\nb^{\perp\nu} \Phi_n
    - \Phi_\nb^{(u)\dagger}\spac\A_n^{\perp\nu} \Phi_\nb \big) \,, \\
   \colored{ P_{13} } &= \colored{ 
    g_{\mu\nu}^\perp\,(i\partial_\perp^\nu Z_v^{\prime\mu})\,
    \big( \Phi_n^\dagger\Phi_\nb + \Phi_\nb^\dagger \Phi_n \big) } \,, \\
   \colored{ P_{14}} &= \colored{ 
    \epsilon_{\mu\nu}^\perp\,(i\partial_\perp^\nu Z_v^{\prime\mu})\,
    \big( \Phi_n^\dagger\Phi_\nb - \Phi_\nb^\dagger\Phi_n \big) } \,.
\end{aligned}
\end{align}
The operators $P_{13}$ and $P_{14}$ can be omitted from the basis if one chooses a reference frame in which the transverse momentum of the $Z'$ boson vanishes. The Wilson coefficients $C_k$ of the operators $P_k$ with $k=5,6,7,8$ are real, while the Wilson coefficients of $P_{11,12}$ satisfy $C_{11,12}^*(u)=C_{11,12}(1-u)$. The remaining Wilson coefficients can be arbitrary complex quantities.

\subsubsection*{Operators containing two or three gauge fields}

A basis of operators in which a $Z'$ boson couples to a pair of gauge fields can be chosen as
\begin{align}\label{Riops}
\begin{aligned}
   R_1 &= Z_v^{\prime\mu}\,\big( 
    \A_{n\mu}^{\perp a}\,i\!\overleftrightarrow{\partial}\!\!_\perp^{\,\nu}\spac \A_{\nb\nu}^{\perp a}
    + \A_{\nb\mu}^{\perp a}\,i\!\overleftrightarrow{\partial}\!\!_\perp^{\,\nu}\spac \A_{n\nu}^{\perp a} 
    \big) \,, \\
   R_2 &= \epsilon_{\nu\alpha}^\perp\,Z_v^{\prime\mu}\,\big( 
    \A_{n\mu}^{\perp a}\,i\!\overleftrightarrow{\partial}\!\!_\perp^{\,\nu}\spac \A_\nb^{\perp\alpha,a} 
    - \A_{\nb\mu}^{\perp a}\,i\!\overleftrightarrow{\partial}\!\!_\perp^{\,\nu}\spac \A_n^{\perp\alpha,a}
    \big) \,, \\
   R_3 &= \epsilon_{\mu\nu}^\perp\,Z_v^{\prime\mu}\,\big( 
    \A_n^{\perp\alpha,a}\,i\!\overleftrightarrow{\partial}\!\!_\perp^{\,\nu}\spac 
    \A_{\nb\alpha}^{\perp a} \big) \,, \\
   R_4 &= \epsilon_{\mu\alpha}^\perp\,Z_v^{\prime\mu}\,\big( 
    \A_n^{\perp\alpha,a}\,i\!\overleftrightarrow{\partial}\!\!_\perp^{\,\nu}\spac \A_{\nb\nu}^{\perp a} 
    - \A_\nb^{\perp\alpha,a}\,i\!\overleftrightarrow{\partial}\!\!_\perp^{\,\nu}\spac \A_{n\nu}^{\perp a} 
    \big) \,, \\
   \colored{ R_5 } &= \colored{ (i\partial_\mu^\perp Z_v^{\prime\mu})\spac\,
    \A_n^{\perp\alpha,a}\,\A_{\nb\alpha}^{\perp a} } \,, \\
   \colored{ R_6 } &= \colored{ (i\partial_\perp^\mu Z_v^{\prime\nu})\,\big(
    \A_{n\mu}^{\perp a}\,\A_{\nb\nu}^{\perp a} + \A_{\nb\mu}^{\perp a}\,\A_{n\nu}^{\perp a} \big) } \,, \\
   \colored{ R_7 } &= \colored{ \epsilon_{\alpha\beta}^\perp\,(i\partial_\mu^\perp Z_v^{\prime\mu})\spac\,
    \A_n^{\perp\alpha,a}\,\A_\nb^{\perp\beta,a} } \,, \\
   \colored{ R_8 } &= \colored{ \epsilon_{\nu\beta}^\perp\,(i\partial_\perp^\mu Z_v^{\prime\nu})\,\big(
    \A_{n\mu}^{\perp a}\,\A_\nb^{\perp\beta,a} - \A_{\nb\mu}^{\perp a}\,\A_n^{\perp\beta,a} \big) } \,, \\
   \colored{ R_9 } &= \colored{ \epsilon_{\mu\alpha}^\perp\,(i\partial_\perp^\mu Z_v^{\prime\nu})\,\big(
    \A_n^{\perp\alpha,a}\,\A_{\nb\nu}^{\perp a} - \A_\nb^{\perp\alpha,a}\,\A_{n\nu}^{\perp a} \big) } \,, \\
   \colored{ R_{10} } &= \colored{ 
    i g_{\mu\nu}^\perp\,Z_v^{\prime\mu} \big( v\cdot\A_n^{\perp a}\,\A_\nb^{\perp\nu,a} 
    + v\cdot\A_\nb^{\perp a}\,\A_n^{\perp\nu,a} \big) = \frac{1}{m_{Z'}}\,\tilde O_{AA}^\parallel } 
    \,, \\[-1mm]
   \colored{ R_{11} } &= \colored{ 
    i\epsilon_{\mu\nu}^\perp\,Z_v^{\prime\mu} \big( v\cdot\A_n^{\perp a}\,\A_\nb^{\perp\nu,a} 
    - v\cdot\A_\nb^{\perp a}\,\A_n^{\perp\nu,a} \big) = \frac{1}{m_{Z'}}\,\tilde O_{AA}^\perp } \,.
\end{aligned}
\end{align}
Because of the relation
\begin{align}
   \epsilon_{\mu\nu}^\perp\,\epsilon_{\alpha\beta}^\perp
   = g_{\mu\alpha}^\perp\,g_{\nu\beta}^\perp - g_{\mu\beta}^\perp\,g_{\nu\alpha}^\perp
\end{align}
it is not necessary to consider operators in which the indices are contracted with two $\epsilon_{\rho\sigma}^\perp$ symbols. The operators $R_5$ to $R_9$ can be omitted from the basis if one chooses a reference frame in which the transverse momentum of the $Z'$ boson vanishes. The operators $R_{10}$ and $R_{11}$, which involve the small components of the gauge fields, can be eliminated using the equations of motion. Generalizing a relation derived in Ref.~\cite{Marcantonini:2008qn} for the case of QCD interactions, we obtain
\begin{align}\label{EOMrela}
   \omega \left( n\cdot\A_n^a \right)_\omega
   &= 2\spac i\partial_\nu^\perp \left( \A_n^{\perp\nu,a} \right)_\omega
    + \int_0^\omega\!d\omega'\,\frac{2\omega'}{\omega}\,
    i f_A^{abc} \left( \A_n^{\perp\nu,b} \right)_{\omega-\omega'}
     \left( \A_{n\nu}^{\perp c} \right)_{\omega'}
    - \frac{2g_A^2}{\omega}\,\sum_\psi \left( \bar\Psi_n\spac\nbsl\,t_A^a\spac\Psi_n \right)_\omega \notag\\
   &\quad\mbox{}- 2g_A^2\,\int_0^\omega\!d\omega'\,\frac{\omega-2\omega'}{\omega}
    \left( \phi_n^\dagger \right)_{\omega-\omega'} t_A^a \left( \phi_n \right)_{\omega'} 
    - 2 g_A^2\,\big( \phi_n^\dagger\,t_A^a\,\phi_0 - \phi_0^\dagger\,t_A^a\,\phi_n \big)_\omega \,.
\end{align}
Here $g_A$ is the relevant gauge coupling, $t_A^a$ are the generators of the gauge group, and the structure constants $f_A^{abc}$ are defined by the Lie algebra $[t_A^a,t_A^b]=if_A^{abc}\,t_A^c$. For the case of $U(1)_Y$, $t_A^a=Y$ is the hypercharge generator and the structure constants vanish. The sum in the third term on the right-hand side runs over all chiral fermion multiplets of the SM that are charged under the corresponding gauge group. The terms involving the Higgs doublet are present for the cases $A=B,W$ only. Following \cite{Marcantonini:2008qn} we use the label formalism, where $\omega$ denotes the eigenvalue of the momentum operator $\overline{\cal P}_n=i\nb\cdot\partial$ acting on $n$-collinear fields. In order to apply this relation to the last two operators in (\ref{Riops}) we use that $v\cdot\A_n^{\perp a}=\frac{v\cdot\nb}{2}\,n\cdot\A_n^{\perp a}$. The operators generated by eliminating the field $n\cdot\A_n^{\perp a}$ using relation (\ref{EOMrela}) are, in the order of appearance, of the form $R_i$ with $i=1,\dots,9$, of the form (\ref{4bosons}) below, of the form $O_3$ and $O_4$, of the form $P_{11}$ and $P_{12}$, and of the form of $P_1$ and $P_2$. The last term on the right-hand side of (\ref{EOMrela}), in particular, implies relation (\ref{eq37}).

Finally, several operators exist in which a $Z'$ boson couples to three gauge fields. They are of the generic form 
\begin{align}\label{4bosons}
\begin{aligned}
   & Z_v^{\prime\mu}\,\A_n^{\perp\nu,a} \A_s^{\alpha,b} \A_\nb^{\perp\beta,c}
    + (n\leftrightarrow\nb) \,, \\
   & Z_v^{\prime\mu}\,\A_n^{(u)\perp\nu,a} \A_n^{\perp\alpha,b} \A_\nb^{\perp\beta,c}
    + (n\leftrightarrow\nb) \,,
\end{aligned}
\end{align}
where the Lorentz and color indices must be contracted in an appropriate way. If the three gauge fields refer to the same gauge group, the index contractions must be performed using the $\epsilon^{abc}$ symbol for $SU(2)_L$ and the $f^{abc}$ or $d^{abc}$ symbols for $SU(3)_c$. Alternatively, one gauge field can be the hypercharge boson and the other two refer to $SU(2)_L$ or $SU(3)_c$, in which case their indices must be contracted using the $\delta^{ab}$ symbol. Note that the operators in the second line carry four transverse Lorentz indices, while for those in the first line the indices $\mu$ and $\alpha$ can be arbitrary.

\renewcommand{\theequation}{B.\arabic{equation}}
\setcounter{equation}{0}

\section{Mass diagonalization in the vector-like quark model}
\label{app:B}

In matrix notation, the mass terms in the vector-like quark model shown in (\ref{eq:Lm}) and (\ref{eq:LagVLQYuk}) can be written in the form
\begin{align}
   \mathcal L_m + \mathcal L_Y
   = - \bar{\mathbb{Q}}_R\,\bm{M}_Q^\dagger \left( \begin{array}{c} \mathbb{Q}_L \\ Q_L \end{array} \right) 
    - \bar{\mathbb{U}}_L\spac\bm{M}_U^\dagger \left( \begin{array}{c} \mathbb{U}_R \\ u_R \end{array} \right)
    - \bar{\mathbb{D}}_L\spac\bm{M}_D^\dagger \left( \begin{array}{c} \mathbb{D}_R \\ d_R \end{array} \right)
    + \text{h.c.} \,,
\end{align}
where $\bm{M}_X^\dagger=(m_X~\,\frac{u}{\sqrt2}\spac\bm{Y}_X^\dagger)$ with $X=Q,U,D$ are $1\times 4$ matrices in all three cases. The squared mass matrices are given by
\begin{align}
\begin{aligned}
   \bm{M}_X\spac\bm{M}_X^\dagger &= \left( \begin{array}{ccc}
    m_X^2 && \frac{u}{\sqrt2}\,m_X \bm{Y}_X^\dagger \\
    \frac{u}{\sqrt2}\,m_X \bm{Y}_X && \frac{u^2}{2}\,\bm{Y}_X \bm{Y}_X^\dagger \end{array} \right) , \\
   \bm{M}_X^\dagger\spac\bm{M}_X &= m_X^2 + \frac{u^2}{2}\,\bm{Y}_X^\dagger \bm{Y}_X 
    \equiv M_X^2 \,.
\end{aligned}
\end{align}
The first one has eigenvalues $M_X$ and 0 (three times), corresponding to the physical mass $M_X$ of the VLQ and the masses of the three light SM quarks in the limit where electroweak symmetry breaking is neglected. The second squared mass matrix is simply a number. We can diagonalize the hermitian matrix $\bm{M}_X\spac\bm{M}_X^\dagger$ by means of the unitary transformation
\begin{align}
   \bm{U}_X^\dagger\spac\bm{M}_X\spac\bm{M}_X^\dagger\spac\bm{U}_X
   = \text{diag}(M_X^2,0,0,0) \,,
\end{align}
where
\begin{align}
   \bm{U}_X = \left( \begin{array}{ccccccc}
    c_X && -s_X && 0 && 0 \\[1mm]
    s_X\spac\hat{\bm{Y}}_X && c_X\spac\hat{\bm{Y}}_X && \bm{n}_X^1 && \bm{n}_X^2 \end{array} \right) .
\end{align}
We have defined the quantities
\begin{align}
   c_X = \frac{m_X}{M_X} \,, \qquad
   s_X = \frac{u}{\sqrt2\spac M_X} \big( \bm{Y}_X^\dagger \bm{Y}_X \big)^{1/2} \,,  \qquad
   \hat{\bm{Y}}_X = \frac{\bm{Y}_X}{\big( \bm{Y}_X^\dagger \bm{Y}_X \big)^{1/2}} \,,
\end{align}
which satisfy $c_X^2+s_X^2=1$ and $\hat{\bm{Y}}_X^\dagger\spac\hat{\bm{Y}}_X=1$. The complex unit vectors $\bm{n}_X^i$ with $i=1,2$ are defined such that $\hat{\bm{Y}}_X^\dagger\spac\bm{n}_X^i=0$ and $\bm{n}_X^{i\spac\dagger}\spac\bm{n}_X^j=\delta^{ij}$. The three unit vectors $\{ \hat{\bm{Y}}_X, \bm{n}_X^1, \bm{n}_X^2 \}$ form an orthonormal basis in generation space. The mass eigenstates of the left-handed $SU(2)_L$ doublets and right-handed $SU(2)_L$ singlets are related to the interaction states by the matrices $\bm{U}_Q^\dagger$ and $\bm{U}_{U,D}^\dagger$, respectively. Written out in components, this gives
\begin{align}
   \left( \mathbb{Q}_L \right)_{\rm mass}  
    = c_Q\,\mathbb{Q}_L + s_Q\spac\hat{\bm{Y}}_Q^\dagger\,Q_L \,, \qquad
   \left( Q_L \right)_{\rm mass}  
    = \left( \begin{array}{c} c_Q\spac\hat{\bm{Y}}_Q^\dagger\,Q_L - s_Q\,\mathbb{Q}_L \\
    \bm{n}_Q^{1\,\dagger}\,Q_L \\ 
    \bm{n}_Q^{2\,\dagger}\,Q_L \end{array} \right) , 
\end{align}
and similarly for the other cases. The second relation, in particular, implies $\big(\hat{\bm{Y}}_Q^\dagger\,Q_L\big)_{\rm mass}=c_Q\spac\hat{\bm{Y}}_Q^\dagger\,Q_L-s_Q\,\mathbb{Q}_L$. It then follows that
\begin{align}
   \mathbb{Q}_L = c_Q \left( \mathbb{Q}_L \right)_{\rm mass}
    - s_Q\spac \big( \hat{\bm{Y}}_Q^\dagger\,Q_L \big)_{\rm mass} \,.
\end{align}
Analogous relations hold for the right-handed components $\mathbb{U}_R$ and $\mathbb{D}_R$ of the $SU(2)_L$-singlet VLQs. 

\end{appendix}

\end{document}